\documentclass[epj,twocolumn, numbook]{svjour}
\usepackage{latexsym}
\usepackage{graphics}
\usepackage{dcolumn}
\usepackage{bm}
\usepackage{graphics}
\usepackage{graphicx}
\usepackage{epsfig}
\usepackage{booktabs,multirow}
\usepackage{hhline}
\usepackage{slashed}
\usepackage{rccol}
\usepackage{mathrsfs,amsmath,amssymb}
\usepackage{xcolor,colortbl}
\usepackage{subfigure}

\usepackage{xcolor}\colorlet{linkequation}{blue}
\usepackage[colorlinks=true, 
            linkcolor=red,   
            filecolor=red,   
            citecolor=blue, 
            urlcolor=blue,
            hyperfootnotes=true]{hyperref}

\usepackage{orcidlink}

\newcommand\ba{\begin{eqnarray}}
\newcommand\ea{\end{eqnarray}}

\begin{document}
\title{Arbitrary $\ell$-state solutions of the Klein-Gordon equation  with the Eckart plus a class of Yukawa potential and its non-relativistic thermal properties}

\author{Mehmet Demirci\,\orcidlink{0000-0003-2504-6251}\inst{1,}\thanks{mehmetdemirci@ktu.edu.tr (Corresponding author) }  \and Ramazan Sever,\orcidlink{0000-0001-6207-8098}\inst{2,}\thanks{sever@metu.edu.tr}
%
}                     
%
\institute{Department of Physics, Karadeniz Technical University, TR61080 Trabzon, Türkiye \and Department of Physics, Middle East Technical University, TR06800 Ankara, Türkiye}
\date{Received: date / Revised version: date}
%

\date{\today}

\abstract{We report bound state solutions of  the Klein Gordon equation with a novel combined potential, the Eckart plus a class of Yukawa potential, by means of the parametric Nikiforov-Uvarov method. To deal the centrifugal and the coulombic behavior terms, we apply the Greene-Aldrich approximation scheme. We present any $\ell$-state energy eigenvalues and the corresponding normalized
wave functions of a mentioned system in a closed form. We discuss various special cases related to our considered potential which are utility for other physical systems and show that these are consistent with previous reports in literature. Moreover, we calculate the non-relativistic thermodynamic quantities (partition function, mean energy, free energy, specific heat and entropy) for the potential model in question, and investigate them for a few diatomic molecules. We find that the energy eigenvalues are sensitive with regard to the quantum numbers $n_r$ and $\ell$ as well as the parameter $\delta$. Our results show that energy eigenvalues are more bounded at either smaller quantum number $\ell$ or smaller parameter $\delta$.
} 

\maketitle

\section{Introduction} 
Nuclei, atoms and molecules, etc., are bombarded with beams of high-energy particles to understand experimentally structure and interactions. As it is well-known, these are called scattering experiments. However, theoretical works are conducted by solving the non-relativistic or the relativistic wave equations for
a known potential. For a quantum system to be accurately described, an analytical solution must be obtained in the form of a wave function that includes all the important properties~\cite{Greiner,Bagrov,Gara,Boivin,Iwo,Mili}.
On other hand, the dynamical behavior of any particle moving at relativistic velocities is described by relativistic quantum mechanics (QM), which present a base in obtaining the energy momentum of fine structure of a hydrogen-like atom. 
Also, for the quantum systems which put the particles into a condition with a stronger potential field, the relativistic effects become important and thus the non-relativistic case needs to be added a correction.
In the relativistic QM, the motion of the scalar particles, spin-0 particles, are described by the Klein-Gordon (KG) equation~\cite{KFG1,KFG2,KFG3,KFG4}. 
Furthermore, the KG equation is suitable for the relativistic particles subject to a general Lorentz scalar and vector potentials. In this regard, the analytical solutions of KG equation for the interaction potential models play an important role in view of relativistic QM.

Many techniques have been developed to solve both non-relativistic and relativistic wave equations with some physical potentials. The following are some of them: Shifted 1/N expansion method \cite{Tang,Roy}, Hartree-Fock method~\cite{Hartree},  perturbation theory~\cite{Stevenson}, the path integral method~\cite{Cai}, factorization~\cite{Dong1}, supersymmetry QM (SUSYQM) \cite{Cooper1,Cooper2,Morales},  Nikiforov-Uvarov (NU) method~\cite{Nikiforov}, and asymptotic iteration method~\cite{aim}. Among them, the NU method has attracted great interest, and its different versions such as the parametric NU method and \cite{ParNU} NU functional analysis (NUFA) method \cite{NUFA} have been developed for this method to be applied easily. By using this technique, many works have been carried out to solve the KG equation with some familiar potentials as follows: Yukawa potential~\cite{Sever2011,Hamzavi2013,Wang2015},  Manning-Rosen Potential~\cite{Jia2013,Wei2010}, Wood-Saxon Potential~\cite{Guo2005,Berkdemir,Badalov},  Hulthén Potential~\cite{Yuan,Ikot2011}, generalized Hulthén potential~\cite{Mehmet,Sever,Qiang}, generalized hyperbolic potential~\cite{Okorie19}, Deng-Fan molecular potentials~\cite{Oluwadre,Ikot21}, inversely quadratic Hellman potential~\cite{Njoku22} and Kratzer Potential~\cite{Qiang2004} and similarly for the case of combined potentials like  Hulthén plus Yukawa potential~\cite{Ahmadov3,Ahmadov21,Demirci21}, Manning–Rosen plus a class of Yukawa~\cite{Demirci20}, Hellmann plus modified Kratzer potential \cite{Aspoukeh}, and Mobius squared plus Eckart potential \cite{Njoku}, etc. 

In regard to enriching the previous attempts, in this study, we propose a novel combined potential model, the Eckart potential plus a class of Yukawa potential, for the first time, in order to calculate the bound state solutions of KG equation. This potential model could be utilized to describe an interaction system which includes the bound and continuum states, and hence can be implemented in the various branches such as atom, molecular, nuclear and particle physics.

The Eckart potential \cite{Eckart30} is defined as
\begin{align}
V_{\text{EP}}(r)=-\frac{V_1 e^{-r/b}}{1-e^{-r/b}}+ \frac{V_2 e^{-r/b}}{(1-e^{-r/b})^2},\,\,\, (V_1,V_2>0),
\label{a1}
\end{align}
with the potential range parameter $b$. Here, $V_1$ and $V_2$ stand for the potential strength parameters. The first part of this potential has a coulomb-like behavior at small values of $r$, while it decreases exponentially for large values of $r$ so that its effect on bound states is smaller compared to the Coulomb potential \cite{Greene}. This potential has a minimum value of $V_{\text{EP}}(r_0) =-\frac{(V_1-V_2)^2}{4V_2}$  at $r_0=a\ln\left(\frac{V_1+V_2}{V_1-V_2}\right)$ for $V_1>V_2$. Also, its second derivative with respect to $r$ at $r = r_0$ leads to the force constant as follows:
\begin{align}
\frac{d^2V_{\text{EP}}}{dr^2}\biggr|_{r=r_0}=\frac{(V_1^2-V_2^2)^2}{8V_1^2V_2^3}.
\label{eq:secder}
\end{align}
Arbitrary $l$-state solutions of the Eckart-type potential have been earlier presented in Refs. \cite{Dong4,Lucha,Liu}.

On the other hand, we also take into account a class of Yukawa potential given as
\ba
V_{\text{CYP}}(r)=-\frac{V_3 e^{-\delta r}}{r}-\frac{V_4 e^{-2\delta r}}{r^2}, \label{potCY}
\ea
with the screening parameter $\delta$. The first part of this potential represents pure the Yukawa potential~\cite{Yukawa} which can be used for explaining the interactions between nucleons.
The potential is monotonically increasing with $r$ and it is negative, implying the force is attractive. In plasma physics, it can be used to describe a charged particle in a weakly non-ideal plasma, and also in electrolytes and colloids, which is also called as the Debye-H\"{u}ckel potential. 
The second part is the inversely quadratic Yukawa potential.

The potentials $V_{\text{EP}}$ and $V_{\text{CYP}}$ have a Coulombic behavior for small values of $r$ but then decrease exponentially when $r$ gets larger. On the other words, they are the screened Coulomb potentials in simple notation. A linear combination of them can be used to examine the deformed-pair interactions of the nucleus and spin-orbit coupling inside the potential fields. An additional fascinating aspect of the combined potential is that it can be utilized to determine the vibrations of the hadronic systems and create a suitable model for other physical phenomena. Based on the above motivations and previous studies, in this work, we propose the following combined potential model, Eckart plus a class of Yukawa potential:
\ba
\begin{split}
V_{\text{ECYP}}(r)&=V_{\text{EP}}(r)+V_{\text{CYP}}(r)\\
&=-\frac{V_1 e^{-r/b}}{1-e^{-r/b}}+ \frac{V_2 e^{-r/b}}{(1-e^{-r/b})^2}-\frac{V_{3} e^{-\delta r}}{r}-\frac{V_{4} e^{-2\delta r}}{r^2}.\label{a4}
\end{split}
\ea
Our aim is to examine it later inside a large quantum system. To this end, we implement the parametric NU method to the problem and use a developed approximation scheme to deal with the coulombic behavior and centrifugal terms. We present the energy eigenvalues and the normalized wave functions for arbitrary $\ell$-states. We also discuss the some special cases by comparing them with the results of the previous works. Furthermore, we provide the thermodynamic quantities of partition function, mean energy, free energy, specific heat and entropy at non-relativistic limit for the potential model in question.

We organize this article in the following:  In Sect.~\ref{br}, we derive the time independent KG equation with the Eckart plus the class of Yukawa potential. In Sect.~\ref{pNU}, we summarize the parametric NU method. In Sect.~\ref{bss}, the bound state solutions of the derived equation are provided via the parametric NU method. In Sec.~\ref{pc}, we present the energy spectrum of some special cases. In Sect.~\ref{thermo}, we provide the thermodynamic properties of the considered system at the nonrelativistic limit. Next, in Sect.~\ref{nr}, we give the numerical results for the bound state solutions and thermodynamic quantities. Finally, in Sect.~\ref{cr}, we summarize our results.

\section{Governing equation}\label{br}

The KG equation consists of two different terms: The scalar rest mass $M$ and the four-vector momentum operator $P_\mu$, hence two types of potential coupling could be included in it. 
The first type is a scalar potential ($V_S$)(with help of exchange $M \to M + V_S$), and the other is a vector potential ($V_V$) (through minimal coupling $P_\mu \to P_\mu-g A_\mu$)~\cite{Greiner}. Accordingly, we have the space-time $V_S$-potentials and the four $V_V$-potentials as $g A_0 = V(t,r)$. In the presence of such potential types, the time-independent KG equation is given as \cite{Demirci20}
\ba
\nabla^{2}\psi+\frac{1}{(\hbar c)^2}\biggl[(E-V_V)^2-(M c^2+V_S)^2 \biggr]\psi=0,
\label{KGequ}
\ea
with the relativistic energy of system $E$. We can rewrite this equation in the natural units, $\hbar = c = 1$, as follows:
\ba
\biggl[-\nabla^{2}+(M+V_S(\textbf{r}))^{2}\biggr]\psi(\textbf{r})=\bigl(E-V_V(\textbf{r})\bigl)^{2}\psi(\textbf{r}).
\label{a5}
\ea
In the presence of a spherical symmetric potential, the wave equation $\psi(\textbf{r}) \equiv\psi(r,\theta,\phi)$ hence can be
separated into angular and radial parts as 
\ba
\psi(r,\theta,\phi)=\frac {\chi(r)}{r}\Theta
(\theta)e^{im\phi},\label{a6}
\ea
where $m \in \mathbb{Z}$ with $m=0,\pm 1,\pm 2,\ldots$. Inserting the result Eq.\eqref{a6} into Eq.\eqref{a5} yields the following radial differential equation:
\ba
\begin{split}
\frac{\text{d}^2\chi(r)}{dr^2}+&\biggl[(E^{2}-M^{2})-2\biggl( V_S(r) M+ V_V(r) E\biggr)+\biggl(V_V^{2}(r)-V_S^{2}(r)\biggr)-\frac{\ell(\ell+1)}{r^{2}}\biggl] \chi(r)=0.
\end{split}
\label{a7}
\ea
In the present work, we assume that $V_S=V_V=V$, and hence the Eq.\eqref{a7} becomes
\ba
\frac{\text{d}^2\chi(r)}{dr^2}+\biggl[(E^{2}-M^{2})-2(M+E) V(r)
-\frac{\ell(\ell+1)}{r^{2}}\biggr] \chi(r)=0. \label{a8}
\ea
It should be noted that the above equation cannot be solved analytically except for $\ell=0$, because of the central term. In this regard, to effectively apply the combined potential~\eqref{a4} to this system, we use the Greene-Aldrich approximation scheme~\cite{Greene,Wen1,Wei,Dong6,Demirci21} as follows:
\ba \frac {1}{r^{2}}\approx
{4\delta^{2}}\frac{e^{-2\delta r}}{(1-e^{-2\delta r})^2},
\label{eq:approx}
\ea
which is a good approximation for $\delta r < < 1$.

We now recall the Eckart potential by setting $\frac{1}{2b}=\delta$ as
\begin{equation} \label{a11}
V'_{\text{EP}}(r)=-\frac{V_1 e^{-2 \delta r}}{1-e^{-2\delta r}}+ \frac{V_2 e^{-2\delta r}}{(1-e^{-2\delta r})^2}.
\end{equation}

Furthermore, the class of Yukawa potential~\eqref{potCY} can be rewritten under the above approximation scheme as
\ba 
V'_{\text{CYP}}(r)=-\frac{2 \delta  V_3 e^{-2\delta r}}{1-e^{-2\delta r}}-\frac{4 \delta^2  V_4 e^{-4\delta r}}{(1-e^{-2\delta r})^2}.
\label{a12} \ea

\begin{figure}[th]
    \begin{center}
\includegraphics[scale=0.42]{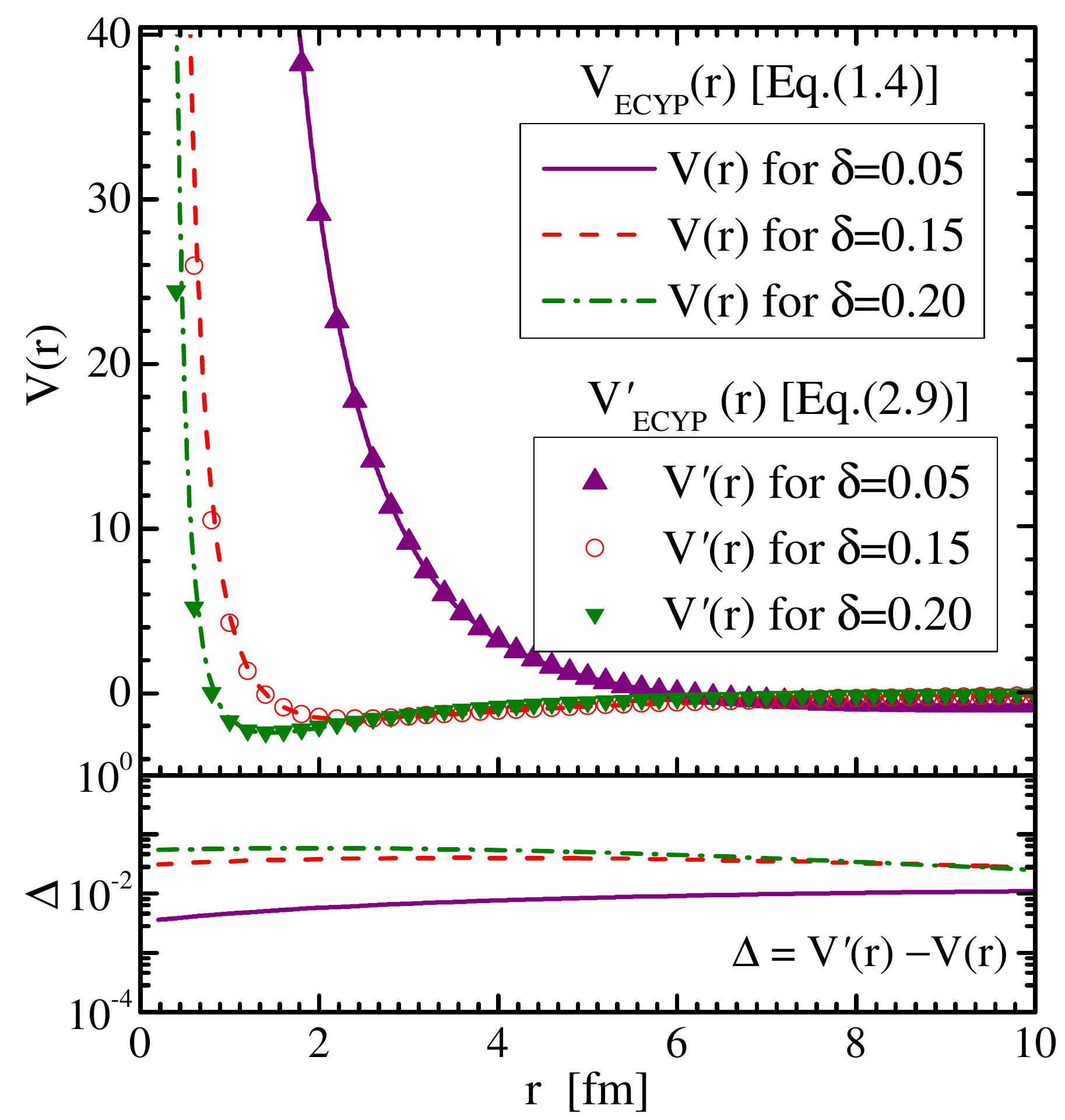}
     \end{center}
   \vspace{-4mm}
\caption{Combined potential and its approximation~(\ref{eq:Vapp}) as a function of $r$ for several values of $\delta$. Here, the potential parameters are set as $V_1=2V_2=V_3=V_4=4$ and $b=1/2\delta$.}
\label{fig1:potapp}
\end{figure}
Consequently, the Eckart plus a class of Yukawa potential becomes
\ba
\begin{split}
V'_{\text{ECYP}}(r)&=V'_{\text{EP}}(r)+V'_{\text{CYP}}(r)\\
&=-\frac{\alpha_1 e^{-2 \delta r}}{1-e^{-2\delta r}}+ \frac{\alpha_2 e^{-2\delta r}}{(1-e^{-2\delta r})^2} +\frac{ \alpha_3 e^{-4\delta r}}{(1-e^{-2\delta r})^2}
\label{eq:Vapp}
\end{split}
\ea
with new definitions of $\alpha_1=(V_{1}+2 \delta  V_3)$, $\alpha_2=V_2$, and $\alpha_3=-4 \delta^2  V_4$. In Fig.~\ref{fig1:potapp}, we also plot the combined potential~\eqref{a4} and its approximation~\eqref{eq:Vapp} as a function of $r$ for various $\delta$. It is obvious from this figure that the approximation becomes more convenient as $\delta$ gets smaller, as expected. 
It implies that Eq.~\eqref{eq:approx} is a good approximation for the small values of $\delta$. In addition, we observe that the combined potential leads to model a system with attractive forces at large distances and repulsive forces at short distances.

Inserting the combined potential~\eqref{eq:Vapp}, the Eckart plus a class of Yukawa potential, into Eq.~\eqref{a8}, then we have
\ba
\begin{split}
\frac{\text{d}^2\chi(r)}{\text{d}r^2}&+\biggl[(E^{2}-M^{2})-2(E+M)\left(-\frac{\alpha_1 e^{-2 \delta r}}{1-e^{-2\delta r}}+ \frac{\alpha_2 e^{-2\delta r}}{(1-e^{-2\delta r})^2} +\frac{ \alpha_3 e^{-4\delta r}}{(1-e^{-2\delta r})^2}\right)\\
&-\frac{4\delta^{2}\ell(\ell+1)e^{-2\delta
r}}{(1-e^{-2\delta r})^2}\biggr] \chi(r)=0, \label{eq:radKG}
\end{split}
\ea
where we define the effective potential as
\ba
V_{\rm eff}(r)= 2(E+M)\left[-\frac{\alpha_1 e^{-2 \delta r}}{1-e^{-2\delta r}}+ \frac{\alpha_2 e^{-2\delta r}}{(1-e^{-2\delta r})^2} +\frac{ \alpha_3 e^{-4\delta r}}{(1-e^{-2\delta r})^2}\right]+\frac{4\delta^{2}\ell(\ell+1)e^{-2\delta
r}}{(1-e^{-2\delta r})^2}.\label{a16}
\ea

\section{Nikiforov-Uvarov Method}\label{pNU}
In this section, we summarize the parametric NU method, which is the new version of the NU method proposed by Tezcan and Sever~\cite{ParNU}. As it is well-known, the NU method is applied to problems which of the differential equation can be converted into the following form of hypergeometric-type equation:
\ba
\frac{\text{d}^{2}\chi(s)}{\text{d}s^{2}}+\frac{\tilde{\tau}}{\sigma}\frac{\text{d}\chi(s)}{\text{d}s}+\frac{\tilde{\sigma}}{\sigma^2}\chi(s)=0.
\label{eq:BasicDif}
\ea

On the other hand, if the considered differential equation for any potential can be written in the generalized form of the Schr\"{o}dinger-like equation,
\begin{eqnarray}
\frac{\text{d}^{2}\chi(s)}{\text{d}s^{2}}+\frac{a_{1}-a_{2}s}{s(1-a_{3}s)}
\frac{\text{d}\chi(s)}{\text{d}s}+\frac{1}{s^2(1-a_{3}s)^{2}} \left(-\xi_{1}s^{2}+\xi_{2}s-\xi_{3}\right)\chi(s)=0, \label{eq:gSl}
\end{eqnarray}
then the parametric NU method, which is a more practical way and detailed below, can be used.
Comparing this with the basic hypergeometric type Eq. \eqref{eq:BasicDif} gives
\begin{eqnarray}
&\tilde{\tau}(s)&=a_{1}-a_{2}s,\\ 
&\sigma(s)&=s(1-a_{3}s),\\
&\tilde{\sigma}(s)&=-\xi_{1}s^{2}+\xi_{2}s-\xi_{3}\,.
\end{eqnarray}
The function $\pi(s)$ becomes
\begin{eqnarray}
\pi(s)=a_{4}+a_{5}s\pm\sqrt{(a_{6}-ka_{3})s^{2}+(a_{7}+k)s+a_{8}}\,,\label{eq:pis}
\end{eqnarray}
with the following parameters
\begin{eqnarray}
\begin{array}{ll}
a_{4}=\frac{1}{2}\,(1-a_{1})\,, & a_{5}=\frac{1}{2}\,(a_{2}-2a_{3})\,, \\
a_{6}=a_{5}^{2}+\xi_{1}\,, &
a_{7}=2a_{4}a_{5}-\xi_{2}\,, \\
a_{8}=a_{4}^{2}+\xi_{3}\,. & \\
\end{array}
\end{eqnarray}
The function under the square root in Eq. \eqref{eq:pis} is to
be the square of a polynomial \cite{Nikiforov}. This condition gives the roots
of the parameter $k$ written as
\begin{eqnarray}
k_{\pm}=-(a_{7}+2a_{3}a_{8})\pm2\sqrt{a_{8}a_{9}}\,,\label{eq:k12}
\end{eqnarray}
where the $k$-values can be imaginary or real, and
$a_{9}=a_{3}a_{7}+a_{3}^{2}a_{8}+a_{6}$\,.
It is obvious that $k_{+}$ and $k_{-}$ in Eq. \eqref{eq:k12} lead to different $\pi(s)$-functions in Eq. \eqref{eq:pis}. For  $k_{-}$, the function $\pi(s)$ becomes
\begin{eqnarray}
\pi(s)=a_{4}+a_{5}s-\left[(a_{3}\sqrt{a_{8}}+\sqrt{a_{9}}\,)s-\sqrt{a_{8}}\,\right]\,,\label{eq:pistb}
\end{eqnarray}
and also we have
\begin{eqnarray}
\tau(s)=a_{1}+2a_{4}-(a_{2}-2a_{5})s-2\left[(\sqrt{a_{9}}
+a_{3}\sqrt{a_{8}}\,)s-\sqrt{a_{8}}\,\right].\label{eq:tau}
\end{eqnarray}
It should be imposed here the following expression for fulfilling the condition that the derivative of the function $\tau(s)$ must be negative in the method
\begin{eqnarray}
\tau^{\prime}(s)&=&-(a_{2}-2a_{5})-2(a_{3}\sqrt{a_{8}}+\sqrt{a_{9}}\,)
\nonumber \\
&=&-2a_{3}-2(a_{3}\sqrt{a_{8}}+\sqrt{a_{9}}\,)<0. \label{eq:tauder}
\end{eqnarray}
In this approach, the energy spectrum equation is calculated from \cite{ParNU}
\begin{eqnarray}
a_{2}n-(2n+1)a_{5}+n(n-1)a_{3}+(2n+1)(a_{3}\sqrt{a_{8}}+\sqrt{a_{9}}\,)
     +a_{7}+2a_{3}a_{8}+2\sqrt{a_{8}a_{9}}=0. \label{eq:eigeneq}
\end{eqnarray}
The weight function $\rho(s)$ from NU-method can be written as
\begin{eqnarray}
\rho(s)=s^{a_{10}-1}(1-a_{3}s)^{\frac{a_{11}}{a_{3}}-a_{10}-1}\,,\label{eq:rho}
\end{eqnarray}
and then we have
\begin{eqnarray}
y_{n}(s)=P_{n}^{(a_{10}-1,\frac{a_{11}}{a_{3}}-a_{10}-1)}(1-2a_{3}s)\,, \label{eq:yns}
\end{eqnarray}
where
\begin{eqnarray}
a_{10}=a_{1}+2a_{4}+2\sqrt{a_{8}}\,\,\text{and}\,
a_{11}=a_{2}-2a_{5}+2(\sqrt{a_{9}}+a_{3}\sqrt{a_{8}})\,.
\end{eqnarray}
The $P_{n}^{(\alpha,\beta)}(1-2a_{3}s)$ are the Jacobi
polynomials. The other part of the general solution is given as
\begin{eqnarray}
\phi(s)=s^{a_{12}}(1-a_{3}s)^{-a_{12}-\frac{a_{13}}{a_{3}}}\,\label{eq:phi}
\end{eqnarray}
with the parameters
\begin{eqnarray}
a_{12}=a_{4}+\sqrt{a_{8}}\,\,\text{and}\,a_{13}=a_{5}-(\sqrt{a_{9}}+a_{3}\sqrt{a_{8}}\,)\,.
\end{eqnarray}

Hence, the general solution $\chi(s)=\phi(s)y(s)$ reads
\begin{eqnarray}
\chi(s)=s^{a_{12}}(1-a_{3}s)^{-a_{12}-\frac{a_{13}}{a_{3}}}
P_{n}^{(a_{10}-1,\frac{a_{11}}{a_{3}}-a_{10}-1)}(1-2a_{3}s)\,.
\end{eqnarray}

\section{Bound-state solutions for the Eckart plus a class of Yukawa potential}\label{bss}
In this section, we solve the bound eigenstates of the KG equation in the presence of our combined potential by means of the parametric NU method.

To this end, we continue from Eq. \eqref{eq:radKG} laid out in Sect. \ref{br}. Inserting the transformation $s=e^{-2\delta r}$ \footnote{It should be noted that $s\in[0,1]$ for $r \in[0,\infty)$ } into Eq. \eqref{eq:radKG}, we get
\ba
\begin{split}
\frac{\text{d}^{2}\chi(s)}{\text{d}s^{2}}+\frac{1-s}{s(1-s)}\frac{\text{d}\chi(s)}{\text{d}s}+\biggl[\frac{1}{s(1-s)}\biggr]^2&\biggl[-\varepsilon^2(1-s)^2-\gamma^{2}s^2-\eta^{2}s +\beta^{2}s(1-s)-\ell(\ell+1)s\biggr]
\chi(s)=0 \label{eq:appdiff}
\end{split}
\ea
with the new parameters
\ba
\varepsilon=\frac{\sqrt{M^2-E^2}}{2\delta}>0~,~~\gamma
=\frac{\sqrt{2(E+M)\alpha_{3}}}{2\delta}>0~,\\
~~\eta
=\frac{\sqrt{2(E+M)\alpha_{2}}}{2\delta}>0~,~~\beta
=\frac{\sqrt{2(E+M)\alpha_{1}}}{2\delta}>0. \label{a19}
\ea
We note that Eq.~\eqref{eq:appdiff} has a suitable form for implementing the parametric NU method. By comparing Eq.\eqref{eq:appdiff} with Eq.\eqref{eq:BasicDif}, we find the expressions as follows
\ba
\begin{split}
\tilde{\tau}(s)&= 1-s,\\
\sigma(s)&=s(1-s),\\
\tilde{\sigma}(s)&=-\varepsilon^2(1-s)^2-\gamma^{2}s^2-\eta^{2}s +\beta^2
s(1-s)-\ell(\ell+1)s. \label{a21}
\end{split}
\ea
We can also rewrite the Eq.~\eqref{eq:appdiff} as
\ba
\begin{split}
\frac{\text{d}^{2}\chi(s)}{\text{d}s^{2}}+\frac{1-s}{s(1-s)}\frac{\text{d}\chi(s)}{\text{d}s}+\biggl[\frac{1}{s(1-s)}\biggr]^2&\biggl[-s^2 (\beta^2+\gamma^2+\varepsilon^2)+s (\beta^2-\eta^2+2\varepsilon^2-\ell(\ell+1)) \\
&-\varepsilon^2 \biggr]
\chi(s)=0.\label{eq:SimEq}
\end{split}
\ea

Comparing Eq.\eqref{eq:SimEq} with \eqref{eq:gSl}, we obtain the parameter set
\begin{eqnarray}
\begin{array}{lll}
\xi_{1}=\beta^2+\gamma^2+\varepsilon^2\,, &\xi_{2}=\beta^2-\eta^2+2\varepsilon^2-l(l+1)\,, & \xi_{3}=\varepsilon^2\,, \\
a_1=a_2=a_3=1\,, &a_4=0\,, &  a_5=-\frac{1}{2}\,,\\
a_6=\frac{1}{4}+\xi_{1}\,,&  a_7=-\xi_{2}\,, & a_8=\xi_{3}\,. \\
a_9=\frac{1}{4}+\xi_{1}-\xi_{2}+\xi_{3}\,,& &   \label{eq:p1}
\end{array}
\end{eqnarray}
From Eq.\eqref{eq:pis}, we obtain the function $\pi(s)$ as
\begin{eqnarray}
\pi(s)=-\frac{1}{2}s\pm\sqrt{\biggr(\frac{1}{4}+\xi_{1}-k\biggr)s^{2}+(k-\xi_{2})s+\xi_{3}}\,.\label{eq:pis2}
\end{eqnarray}

Substituting \eqref{eq:p1} into \eqref{eq:k12}, we obtain its roots as
\begin{eqnarray}
k_{\pm}=-(-\xi_{2}+2\xi_{3})\pm2\sqrt{\xi_{3} \bigl(\frac{1}{4}+\xi_{1}-\xi_{2}+\xi_{3}\bigr)}\,.\label{eq:k121}
\end{eqnarray}
For $k_{-}$, we find the convenient functions $\pi(s)$ and $\tau(s)$ from Eq.\eqref{eq:pistb} and Eq.\eqref{eq:tau}, respectively as 
\begin{eqnarray}
\pi(s)=-\frac{1}{2}s-\left[\biggl(\sqrt{\frac{1}{4}+\xi_{1}-\xi_{2}+\xi_{3}}+\sqrt{\xi_{3}}\,\biggr)s-\sqrt{\xi_{3}}\,\right]\,,\label{eq:pistb2}
\end{eqnarray}
and
\begin{eqnarray}
\tau(s)=1-2s-2\left[\biggl(\sqrt{\frac{1}{4}+\xi_{1}-\xi_{2}+\xi_{3}}+\sqrt{\xi_{3}}\,\biggr)s-\sqrt{\xi_{3}}\,\right].\label{eq:tau2}
\end{eqnarray}

The derivative of the function $\tau(s)$ from Eq.\eqref{eq:tauder} can be obtained as
\begin{eqnarray}
\tau^{\prime}(s)=-2-2\biggl(\sqrt{\frac{1}{4}+\xi_{1}-\xi_{2}+\xi_{3}}+\sqrt{\xi_{3}}\,\biggr)<0, \label{eq:tauder2}
\end{eqnarray}
which is the essential condition for bound-state real solution.
In this approach, substituting \eqref{eq:p1} into Eq.\eqref{eq:eigeneq}, we obtain the energy spectrum equation for the Eckart plus a class of Yukawa potential as
\ba
M^2-E_{n_{r},\ell}^{2}=\left[\frac{\beta^{2}-\eta^{2}-\ell(\ell+1)-1/2-n_r(n_r+1)-(2n_{r}+1)\sqrt{\frac{1}{4}+\gamma^2+\eta^{2}+\ell(\ell+1)}}{n_r+\frac{1}{2}+\sqrt{\frac{1}{4}+\gamma^2+\eta^{2}+\ell(\ell+1)}}\times\delta\right]^2~~~~
\label{a34}
\ea
or in more compact form as
\ba M^2-E_{n_{r},\ell}^2 =4\delta^2 \left[
\frac{\beta^2+\gamma^2-(n_r+\omega)^2
}{ 2(n_r+\omega)}\right]^2, \label{eq:Enr2} 
\ea
where $\omega=\frac{1}{2}+\sqrt{\frac{1}{4}+\gamma^2+\eta^{2}+\ell(\ell+1)}$.
We can calculate numerically the energy eigenvalues from the above relation. 

We now move on derivation of the radial eigenfunctions. First, we write the weight function  $\rho(s)$ from Eq.\eqref{eq:rho} as
\ba
\rho(s)=s^{2\varepsilon}(1-s)^{2\omega-1} \label{eq:rho2}
\ea
with $a_{10}=1+2\varepsilon\, $ and $ a_{11}=2+2\biggl(\sqrt{\frac{1}{4}+\xi_{1}-\xi_{2}+\xi_{3}}+\varepsilon\biggr)\,$. In order to find the exact solution, we set the wave function as $\chi(s)=y(s)\phi(s)$. Thus, for the first part of wave function, we have 
\ba
y_{n_{r}}(s) = C_{n_{r}} P_{n_{r}}^{(2\varepsilon,2\omega-1)}(1-2s)
\label{eq:yn2}
\ea
from Eq.~\eqref{eq:yns}. We obtain the other part of wave function from Eq. \eqref{eq:phi} as
\ba \phi (s)=s^{\varepsilon}(1-s)^\omega, \label{eq:phi2}
\ea
with $a_{12}=\sqrt{\xi_3}=\varepsilon\,$ and $\,a_{13}=-\frac{1}{2}-\sqrt{\frac{1}{4}+\xi_{1}-\xi_{2}+\xi_{3}}-\varepsilon$.

Then, inserting Eqs.~\eqref{eq:yn2} and~\eqref{eq:phi2} into the
general solution gives
\ba
\chi_{n_{r}}(s)=C_{n_{r}}s^{\varepsilon}(1-s)^\omega
P_{n_{r}}^{(2\varepsilon,2\omega-1)}(1-2s). \label{a48}
\ea

Using the following relation \cite{Abramowitz}
\ba
P_n^{(a,b)}(1-2s)=\frac{{\Gamma(n+a+1)}}{{n!\Gamma(a+1)}}
{}_2F_{1} \left({-n,1+n+a+b,1+a;s}\right), \label{a49}
\ea
we can write Eq.~\eqref{a48} as
\ba \chi_{n_{r}}
(s)=C_{n_{r}}s^{\varepsilon}(1-s)^{\omega}\frac{\Gamma (n_{r}+2\varepsilon+1)}{n_{r}!\Gamma (2\varepsilon+1)} {}_2F_{1}
\left({-n_{r},n_{r}+{2\varepsilon}+2\omega,1+{2\varepsilon};s}\right).
\label{a50}
\ea
where we can determine the normalization constant $C_{n_{r}}$ by 
\ba \int\limits_0^\infty
|R(r)|^{2}r^{2}dr=\int\limits_0^{\infty} |\chi(r)|^{2}
dr=\frac{1}{2\delta}\int\limits_0^1\frac{1}{s}|\chi (s)|^{2}ds=1.
\label{a51}
\ea
With help of the following identity
\cite{Abramowitz}
\ba
\begin{split}
\int\limits_0^1 {(1-z)^{2(\nu+1)}z^{{2\lambda}-1}}&
\biggl[{ {}_2F_{1} (-n,2(\nu+\lambda+1)+n,2\lambda+1;z)} \biggr]^2 dz \\
&= \frac{{n!(n+{\nu}+1)\Gamma (2\lambda)\Gamma (n+{2\nu}+2)\Gamma({2\lambda}+1)}}{{(n+{\nu}+{\lambda}+1)\Gamma(n+{2\lambda}+1)\Gamma (2({\nu}+{\lambda}+1)+n)}},
 \label{a52}
\end{split}
\ea
where $\nu > \frac{{-3}}{2}$ and $\lambda>0 $, we calculate the normalization constant as
\ba C_{n_r}=\sqrt{\frac{2 n_{r}! \delta (n_{r}+\omega+\varepsilon)\Gamma(n_r+{2\varepsilon}+2\omega)\Gamma
({2\varepsilon}+1)}{(n_{r}+\omega)\Gamma
(2\varepsilon)\Gamma(n_r+{2\varepsilon}+1) \Gamma (n_{r}+2\omega)}}. \label{a53} \ea

Hence, we can write the total wave function $\psi(r,\theta,\phi)$ for the Eckart plus a class of Yukawa potential as
\ba 
\psi(r,\theta,\phi)=N_{n_{r}l}\frac{1}{r} (e^{-r/2b})^{\varepsilon}(1-e^{-r/2b})^{\omega} {}_2F_{1}
\left({-n_{r},{2\varepsilon}+2\omega+n_{r},1+{2\varepsilon};e^{-r/2b}}\right) Y_{lm}(\theta,\phi),
\label{eq:fullwave}
\ea
where $Y_{lm}(\theta,\phi)$ is the spherical harmonics.

\section{Particular cases}\label{pc}
In this section, we discuss some particular cases. These are several well-known potentials that are useful for other physical systems. We construct them by adjusting our potential parameters.
\begin{enumerate}
  \item Taking $V_{3}=V_{4}=0$, the total potential reduces to \textit{the central Eckart potential}. Hence, we obtain the energy spectrum equation of the new case as
\ba 
M^2-E_{n_{r},\ell}^2 =4\delta^2 \left[
\frac{\frac{(E+M)}{2\delta^2}V_{1}-(n_r+\omega)^2
}{ 2(n_r+\omega)}\right]^2, \label{eq:Enr_Eckart} 
\ea
where $\omega=\frac{1}{2}+\sqrt{\frac{1}{4}+\ell(\ell+1)+\eta^2}$.
This result is the same with expression (16) of Ref.~\cite{zang08} under exchange $ V_1 \leftrightarrow V_2 $. Also, for the s-wave case ($l=0$), this is consistent with that given in Eq. (15) of Ref.~\cite{Liu}.
Also, the corresponding wave function can be written as
\ba
\chi_{n_{r},\ell}(s)=C_{n_{r},\ell}s^{\frac{\sqrt{M^2-E^2}}{2\delta}}(1-s)^{\frac{1}{2}+\sqrt{\frac{1}{4}+\ell(\ell+1)+\eta^2}}
P_{n_{r}}^{\bigl(\frac{\sqrt{M^2-E^2}}{\delta},2\sqrt{\frac{1}{4}+\ell(\ell+1)+\eta^2}\bigr)}(1-2s).
\ea

\item Furthermore, it is known that the Eckart potential can be reduced to \textit{the Hulthén potential}. This is achieved by taking $V_2= V_3=V_4=0$. If so, we have from  Eq.\eqref{eq:Enr2} 

\ba M^2-E_{n_{r},\ell}^2 =4\delta^2 \left[
\frac{\frac{(E+M)}{2\delta^2}V_{1}-(n_r+\ell+1)^2
}{ 2(n_r+\ell+1)}
\right]^2. \label{eq:hultenEnr2} 
\ea
This is the same as the previously obtained expression in Eq.(50) of Ref. \cite{Ahmadov3} by selecting $S(r) = V (r)$. The corresponding wave function can be written as
\ba
\chi_{n_{r},\ell}(s)=C_{n_{r},\ell}s^{\frac{\sqrt{M^2-E^2}}{2\delta}}(1-s)^{1+\ell}
P_{n_{r}}^{\bigl(\frac{\sqrt{M^2-E^2}}{\delta},1+2\ell\bigr)}(1-2s).
\ea

\item Setting $V_{2}$ and $V_{4}$ to zero leads to the case of \textit{the Hulthén plus a class of Yukawa potential}. If so, we have from  Eq.\eqref{eq:Enr2} 

\ba M^2-E_{n_{r},\ell}^2 =4\delta^2 \left[
\frac{\beta^2-(n_r+\ell+1)^2
}{ 2(n_r+\ell+1)}
\right]^2 . \label{eq:hultenEnr2} 
\ea
This result is consistent with the expressions obtained in Eq.(47) of Ref. \cite{Ahmadov3} and Eq.(42) of Ref. \cite{Ahmadov21}. We present the corresponding wave function as
\ba
\chi_{n_{r},\ell}(s)=C_{n_{r},\ell}s^{\frac{\sqrt{M^2-E^2}}{2\delta}}(1-s)^{1+\ell}
P_{n_{r}}^{\bigl(\frac{\sqrt{M^2-E^2}}{\delta},1+2\ell\bigr)}(1-2s).
\ea

\item Setting $V_{1}$ and $V_{2}$ to zero leads to the case of \textit{a class of Yukawa potential} \eqref{potCY}. If so, we have the following equation  for energy eigenvalues:
\ba
M^{2}-E^{2}=\left[\frac{\xi^{2}-1/2-\ell(\ell+1)-n_r(n_r+1)-(2n_{r}+1)\sqrt{\frac{1}{4}+\ell(\ell+1)+\zeta^{2}}}{n_r+\frac{1}{2}
+\sqrt{\frac{1}{4}+\ell(\ell+1)+\zeta^{2}}}\right]^2\delta^2, \label{aclasYuk}
\ea
with parameters
\ba
\xi=\frac{\sqrt{4\delta V_{3}(E+M)}}{2\delta}, \label{aclasYukpar1}
\ea
and
\ba
\zeta=\sqrt{-2 V_{4}(E+M)}. \label{aclasYukpar2}
\ea
The associated wave function can be written as
\ba
\chi_{n_{r},\ell}(s)=C_{n_{r},l}s^{\frac{\sqrt{M^2-E^2}}{2\delta}}(1-s)^{\frac{1}{2}+\sqrt{\frac{1}{4}+\ell(\ell+1)+\zeta^2}}
P_{n_{r}}^{\bigl(\frac{\sqrt{M^2-E^2}}{\delta},2\sqrt{\frac{1}{4}+\ell(\ell+1)+\zeta^2}\bigr)}(s).
\ea
These results are the same with the relations obtained in Eqs.(77) and (78) of Ref.~\cite{Demirci20}.

\item By limiting $\delta \to 0$, the class of Yukawa potential can be approximated as
\ba
\begin{split}
V_{\text{KFP}}&=\lim_{\delta \to 0}\left(-\frac{V_3 e^{-\delta r}}{r}-\frac{V_4 e^{-2\delta r}}{r^2}\right) \\ &\simeq -\frac{V_3 }{r}-\frac{V_4 }{r^2}.
\end{split}
\ea
Setting $V_3=2r_e D_e$ and $V_4=-r_e^2 D_e$ in above expression, then we have \textit{the Kratzer–Fues potential}. The parameters $D_e$ and $r_e$ stand for the separation energy and the equilibrium distance between two bodies, respectively. We obtain its energy spectrum as
\ba
\begin{split}
M^2&-E^{2}=\frac{(2r_e D_e)^2 (E+M)^2 }{\ell(\ell+1)+n_r(n_r+1)+\frac{1}{2}+2 r_e^2 D_e (E+M)+\left(n_r+\frac{1}{2}\right) \sqrt{(2 \ell+1)^2+8 r_e^2 D_e (E+M)}}.
\end{split}
\ea
  \item Taking $V_{1}=V_{2}=V_{4}=0$ in the combined potential, we obtain \textit{the central Yukawa potential}. Then, we can write its energy eigenvalue equation as
\ba
M^{2}-E^{2}=\left[\frac{\xi^{2}-1/2-\ell(\ell+1)-n_r(n_r+1)-(2n_{r}+1)(\ell+\frac{1}{2})}{n_r+\ell+1}\right]^2 \delta^2, \label{a76}
\ea
where $\xi$ is taken as in Eq.~\eqref{aclasYukpar1}.
This is consistent with those in Ref.~\cite{Wang2015} for the constant mass case. One can easily see this by taking $\alpha\rightarrow\delta$ and $q=1$ in Eq.(39) of Ref.~\cite{Wang2015}.
We get its eigenfunctions as
\ba
\chi_{n_{r},\ell}(s)=C_{n_{r},\ell}s^{\frac{\sqrt{M^2-E^2}}{2\delta}}(1-s)^{\frac{1}{2}+\sqrt{\frac{1}{4}+\ell(\ell+1)}}
P_{n_{r}}^{\bigl(\frac{\sqrt{M^2-E^2}}{\delta},2\sqrt{\frac{1}{4}+\ell(\ell+1)}\bigr)}(s).
\ea
  \item Setting $V_{1}=V_{2}=V_{3}=0$, we get \textit{the inversely quadratic Yukawa potential}. If so, we have the energy spectrum equation as
\ba
M^{2}-E^{2}=\left[\frac{-1/2-\ell(\ell+1)-n_r(n_r+1)-(2n_{r}+1)\sqrt{\frac{1}{4}+\zeta^{2}+\ell(\ell+1)}}{n_r+\frac{1}{2}+\sqrt{\frac{1}{4}+\zeta^{2}+\ell(\ell+1)}}\right]^2\delta^2 \label{a762}
\ea
with $\zeta$ presented as in Eq.~\eqref{aclasYukpar2}.
We also get the associated eigenfunctions as
\ba
\chi_{n_{r},\ell}(s)=C_{n_{r},\ell}s^{\frac{\sqrt{M^2-E^2}}{2\delta}}(1-s)^{\frac{1}{2}+\sqrt{\frac{1}{4}+\zeta^{2}+\ell(\ell+1)}}
P_{n_{r}}^{\bigl(\frac{\sqrt{M^2-E^2}}{\delta},2\sqrt{\frac{1}{4}+\zeta^{2}+\ell(\ell+1)}\bigr)}(s).
\ea
These results are the same with Eqs.(84) and (85) of Ref. \cite{Demirci20}.
  \item  If we take $\delta\rightarrow 0$ in the case of $V_{1}=V_{2}=V_{4}=0$,  the combined potential turns to  \textit{the Coulomb-like potential}, $V(r)=-V_{3}/r$, and so we have 
\ba
M^{2}-E^{2}=\left[\frac{ V_3 (E+M)}{(1+\ell+n)}\right]^2\label{eq:EnerCol}.
\ea
Then, from Eq.\eqref{eq:EnerCol}, we obtain
\ba
E=M\frac{(1+n+\ell)^2- V_3^2 }{(1+n+\ell)^2+ V_3^2}\label{eq:E_Clp},
\ea
which is consistent with Eq.(51) of Ref.~\cite{Wang2015} and those
results in Ref.~\cite{Ikhdair09}.
  \item For the \textit{s-wave case} ($\ell=0$), the centrifugal term disappears from Eq.\eqref{eq:radKG}. As a result, we have the following energy spectrum equation 
\ba M^2-E_{n_{r},l}^2 =4\delta^2 \left[
\frac{\beta^2+\gamma^2-\biggl(n_r+\frac{1}{2}+\sqrt{\frac{1}{4}+\gamma^2+\eta^{2}}\biggr)^2
}{ 2\biggl(n_r+\frac{1}{2}+\sqrt{\frac{1}{4}+\gamma^2+\eta^{2}}\biggr)}\right]^2. \label{eq:Enr2l0} 
\ea

\end{enumerate}

\section{Thermodynamic quantities at the nonrelativistic limit}\label{thermo}
In this section, we will examine the thermodynamic properties of the Eckart plus a class of Yukawa potential at the non-relativistic limit. Now mapping $E_{nl}-M\rightarrow E_{nl}$  and $E_{nl}+M\rightarrow \frac{2\mu}{\hslash^2}$ in Eq.~\eqref{eq:Enr2}, we derive the non-relativistic energy spectrum for the potential model in question
\ba  
E_{n,l}=-\frac{2 \delta^2 \hslash^2}{\mu} \left[\frac{\kappa}{ 2(n+\nu)}- \frac{(n+\nu)}{2}\right]^2,
\label{eq:Enr_nonrel} 
\ea
where  
\ba
\kappa&=&\frac{\mu}{\delta^2 \hslash^2}(\alpha_1+\alpha_3), \\
\nu &=&\frac{1}{2}+\sqrt{\frac{1}{4}+\frac{\mu}{\delta^2 \hslash^2}(\alpha_2+\alpha_3)+l(l+1)}.
\ea
To determine the thermal properties of the system, we first need to derive its partition function. In statistical physics, the partition function, as a function of temperature, is often considered a distribution function, and once known, other thermal properties such as internal energy, entropy, free energy and specific heat capacity  can be derived from it. These quantities can either be calculated theoretically or experimentally.
\begin{description}
  \item[i. \textit{Partition function}:]
The partition function can be calculated by summation over all possible energy levels at a given temperature $T$ and is defined as \cite{Jia17}
\ba
Z(\beta)=\sum_{n=0}^{\lambda} e^{-\beta E_{nl}},\label{Eq:PartFunc}
\ea
where $\beta=1/(k_B T)$, $k_B$ is the Boltzmann constant and $\lambda$ is the upper bound vibration quantum number. We obtain the parameter $\lambda$ as
\ba
{\frac{\text{d} E_{n,l}}{\text{d}n}}\biggr|_{n=\lambda}=0\Longrightarrow \lambda=\sqrt{\kappa}-\nu. \label{Eq:lambda}
\ea

We use the Poisson summation formula for a finite summation with the upper bound to evaluate the partition function in Eq. \eqref{Eq:PartFunc}, given by \cite{Morse53,Strekalov7,Song17} 
\ba
\sum_{n=0}^{N} f(n)=\frac{1}{2} \biggl[f(0)-f(N+1) \biggr]+ \int_0^{N+1} f(x) \text{d}x, \label{Eq:PartFunc2}
\ea
where $N$ is taken as $\lambda$ rounded to an integer. With help of this formula, we can write the partition function as
\ba
Z(\beta)=\frac{1}{2} \biggl[(e^{A \beta  \left(\frac{\kappa }{2 \nu }-\frac{\nu }{2}\right)^2}-e^{ A \beta  \left(\frac{\kappa }{2 (\nu +N+1)}-\frac{\nu +N+1}{2} \right)^2 } \biggr]+ \int_0^{N+1} e^{ A \beta  \left(\frac{\kappa }{2 (\nu +x)}-\frac{\nu +x}{2} \right)^2 } \text{d}x,\label{Eq:PartFunc3}
\ea
where $ A= \frac{2 \delta^2 \hslash^2}{\mu}$.
Then, we calculate it as 
\ba
\begin{split}
Z(\beta)=&\frac{e^{-A \beta \kappa }}{2 \sqrt{A \beta}} \biggl[ \sqrt{A\beta } \biggl(e^{\frac{A \beta  \left(\kappa +\nu ^2\right)^2}{4 \nu ^2}}-e^{\frac{1}{4} A \beta  \left(4 \kappa +\left(\nu+N+1-\frac{\kappa }{\nu +N+1}\right)^2\right)}\biggr)\\
&-\sqrt{\pi } \biggl( Y_1(\nu)-Y_1(\nu +N+1)\biggr) +\sqrt{\pi } e^{A \beta  \kappa } \biggl( Y_2(\nu) -Y_2(\nu +N+1)\biggr) \biggr] \label{Eq:PartFunc0}
\end{split}
\ea
with the following definitions
\ba
Y_1(x)=\emph{Erfi}\left(\frac{\sqrt{A\beta} \left(\kappa +x^2\right)}{2 x }\right), \\
Y_2(x)=\emph{Erfi}\left(\frac{\sqrt{A\beta} \left(\kappa -x^2\right)}{2 x }\right),
\ea
where the imaginary error function $\emph{Erfi}$ is given by \cite{Abramowitz}
\ba
\emph{Erfi}(z)=-i\emph{Erf}(iz)=\sqrt{\frac{4}{\pi}} \int_0^z e^{u^2} \text{d}u.
\ea

Using the partition function \eqref{Eq:PartFunc0}, we derive the other thermodynamic quantities in the following items.
  \item[ii. \textit{Mean energy}:] It is defined as the energy included in a thermodynamic system and it is necessary to prepare or improve the system in its current internal state. For an isolated system, it is constant. We obtain the mean energy as
  \ba 
  U(\beta)=-\frac{\partial }{\partial \beta}\,\ln Z(\beta)=\frac{\Lambda_1}{\Lambda_2}
  \ea
  with
\ba
\begin{split}
  \Lambda_1=&-2 (A \beta)^{3/2} \biggl[ (B_1-\kappa ) e^{A  B_1 \beta }-(B_2-\kappa ) e^{A B_2 \beta }\biggr]+2 \sqrt{A \beta } \biggl[ f_1(\nu ) e^{A \beta  f_1(\nu )^2}-f_1(\nu +N+1) e^{A \beta  f_1(\nu +N+1)^2} \\
  &-f_2(\nu ) e^{A \beta  \left(f_2(\nu )^2+\kappa \right)}+f_2(\nu +N+1) e^{A \beta  \left(f_2(\nu +N+1)^2+\kappa \right)}\biggr] + \sqrt{\pi } \biggl[ e^{A \beta  \kappa } \biggl(Y_2(\nu )-Y_2(\nu +N+1)\biggr)\\
  &-(2 A \beta  \kappa +1) \biggl( Y_1(\nu )-Y_1(\nu +N+1) \biggr)\biggr]\\
\Lambda_2=&2 \sqrt{A} \beta^{3/2} \left(e^{A B_1 \beta }-e^{A B_2 \beta}\right)-2 \sqrt{\pi } \beta  \biggl[Y_1(\nu )-Y_1(\nu +N+1)+ e^{A \beta  \kappa } \biggl(Y_2(\nu +N+1)-Y_2(\nu ) \biggr) \biggr]
\end{split}
\end{eqnarray}
where
  \begin{eqnarray}
  \begin{split}
&B_1=\frac{\left(\kappa +\nu ^2\right)^2}{4 \nu ^2}, \,\,
B_2=\frac{1}{4} \left(4 \kappa +\left(\nu+N+1 -\frac{\kappa }{\nu +N+1}\right)^2\right),\\
&f_1(\nu)=\frac{\kappa +\nu ^2}{2 \nu }, \,\,f_2(\nu)=\frac{\kappa -\nu ^2}{2 \nu }.
\end{split}
  \end{eqnarray}
  \item[iii. \textit{Free energy}:] This is a thermodynamic potential which provides a forecast of the helpful work obtained from a closed system with a constant temperature. We calculate the free energy from 
  \ba F(\beta)=-\frac{1}{\beta}\ln Z(\beta). \ea
  
  \item[iv. \textit{Specific heat capacity}:] In thermodynamics, the specific heat capacity is also named as massic heat capacity that is the heat capacity of a sample of the substance divided by its mass.  We calculate the specific heat quantity with \footnote{
It is not explicitly presented here as it has a rather long analytical expression; however, its graphical representation is presented in Sect. \ref{nr}.  
}
  \ba
\begin{split}    
   C(\beta)&=-k_B \beta^2\frac{\partial U(\beta)}{\partial \beta}\\ &=k_B \beta^2\frac{\partial^2 \ln Z(\beta)}{\partial \beta^2}.
  \end{split}
  \ea
  \item[v. \textit{Entropy}:] It is defined as the measure of the amount of thermal energy per unit temperature of a system that cannot be used to provide any productive work. Also, we can think of the amount of entropy as a measure of the unpredictability or disorder of the system. In our case, we derive the entropy as
  \ba
\begin{split}  
  S(\beta)=k_B \beta^2\frac{\partial F(\beta)}{\partial \beta}&=k_B\ln Z(\beta)-k_B\beta \frac{\partial \ln Z(\beta)}{\partial \beta}\\&=k_B\ln Z(\beta)+k_B\beta \frac{\Lambda_1}{\Lambda_2}.
\end{split}
\ea
\end{description}

\section{Numerical evaluation}\label{nr}
\begin{figure}[h]
    \begin{center}
\includegraphics[scale=0.45]{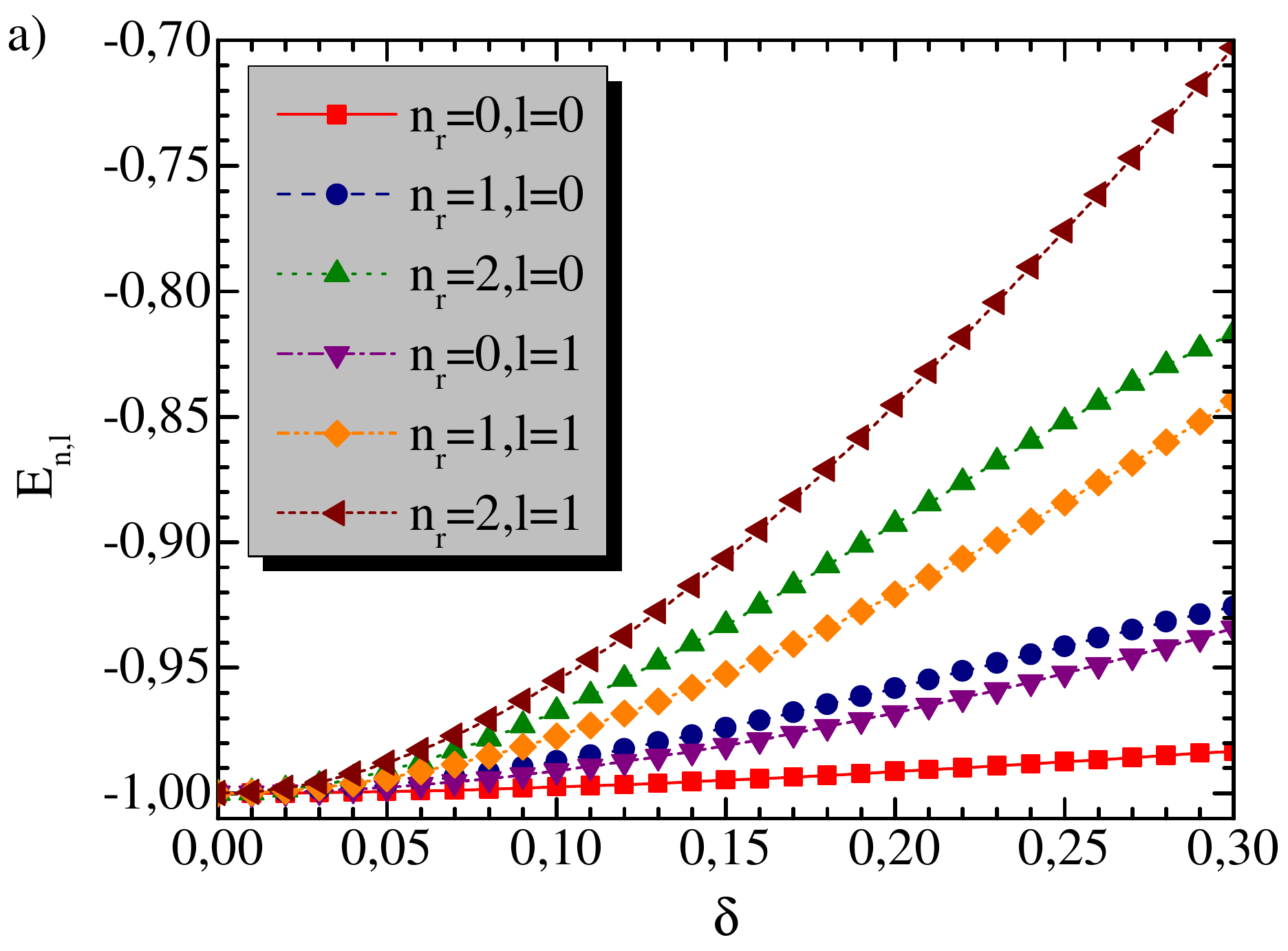}
\includegraphics[scale=0.45]{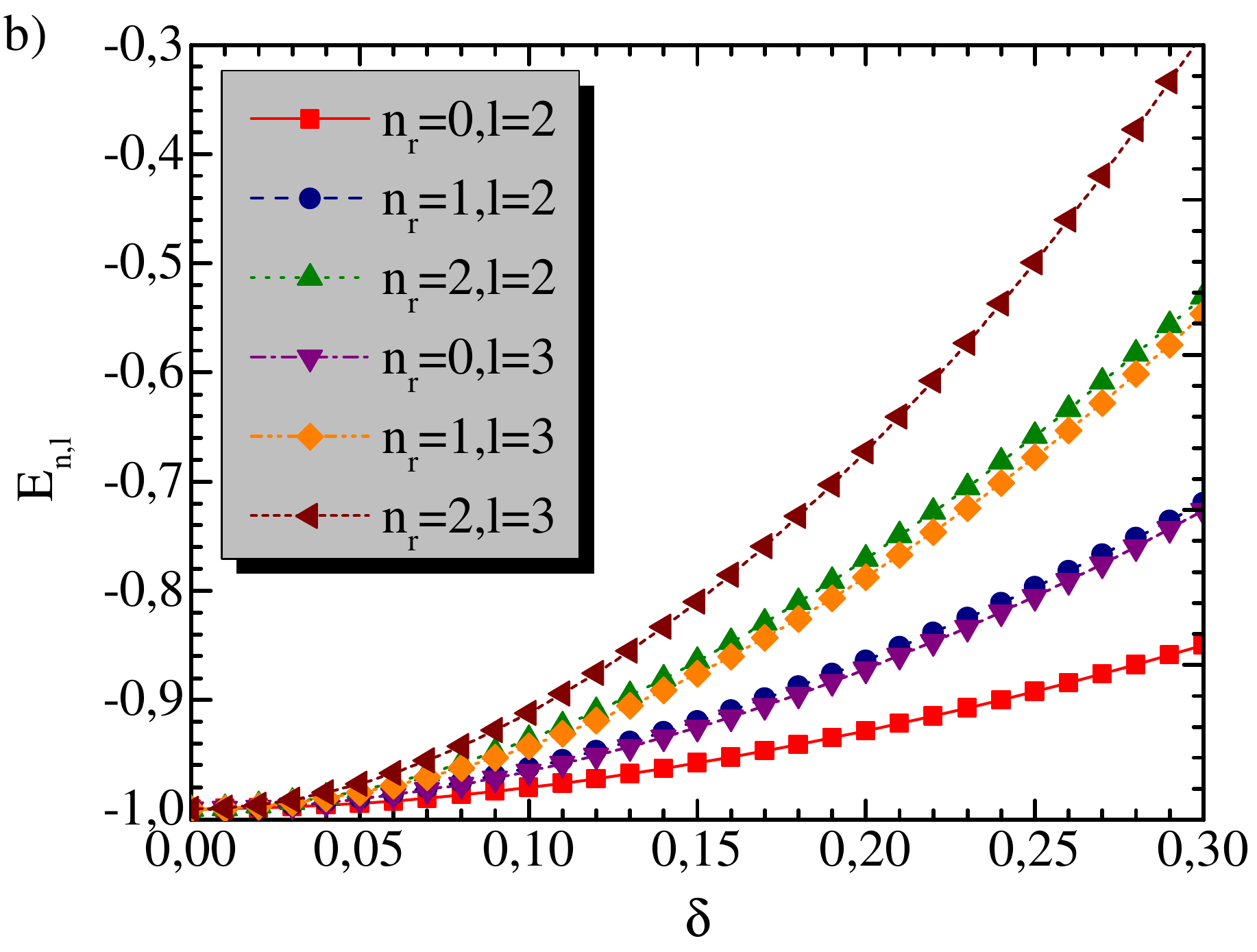}
     \end{center}
          \vspace{-2mm}
\caption{ Energy levels of $E_{n_r,\ell}$ as a function of $\delta$ for given $n_r=0,1,2$ with a) $\ell =0,1$ and b) $\ell =2,3$.\label{fig:Edelta}}
\end{figure}

Now, we discuss the numerical results for the bound state solutions and the non-relativistic thermal properties of the Eckart plus a class of Yukawa potential. First, we examine the energy levels E as a function of the screening parameter $\delta$ and quantum number $n_r$ for arbitrary $\ell$. For simplicity, we set the free parameters as follows: $V_1=2V_2=V_3=V_4=4$, $M=1$, and $b=1/2\delta$, unless otherwise stated. We also consider the natural units here ($\hbar=c=1$).
We show the energy eigenvalues in Fig.~\ref{fig:Edelta} as a function of $\delta$ for $n_r=0,1,2$ and $\ell=0,1,2,3$. Here, the $\delta$ ranges from 0 to 0.30. It is clear that the energy eigenvalues increase as the $\delta$ increases. The energy levels with the same total value of $n_r + \ell$ have close values to each other. For example, $(n_r,\ell) = (0, 1), (1, 0)$, $(n_r,\ell) = (0, 2), (1, 1), (2, 0)$. 

\begin{figure}[h]
    \begin{center}
\includegraphics[scale=0.45]{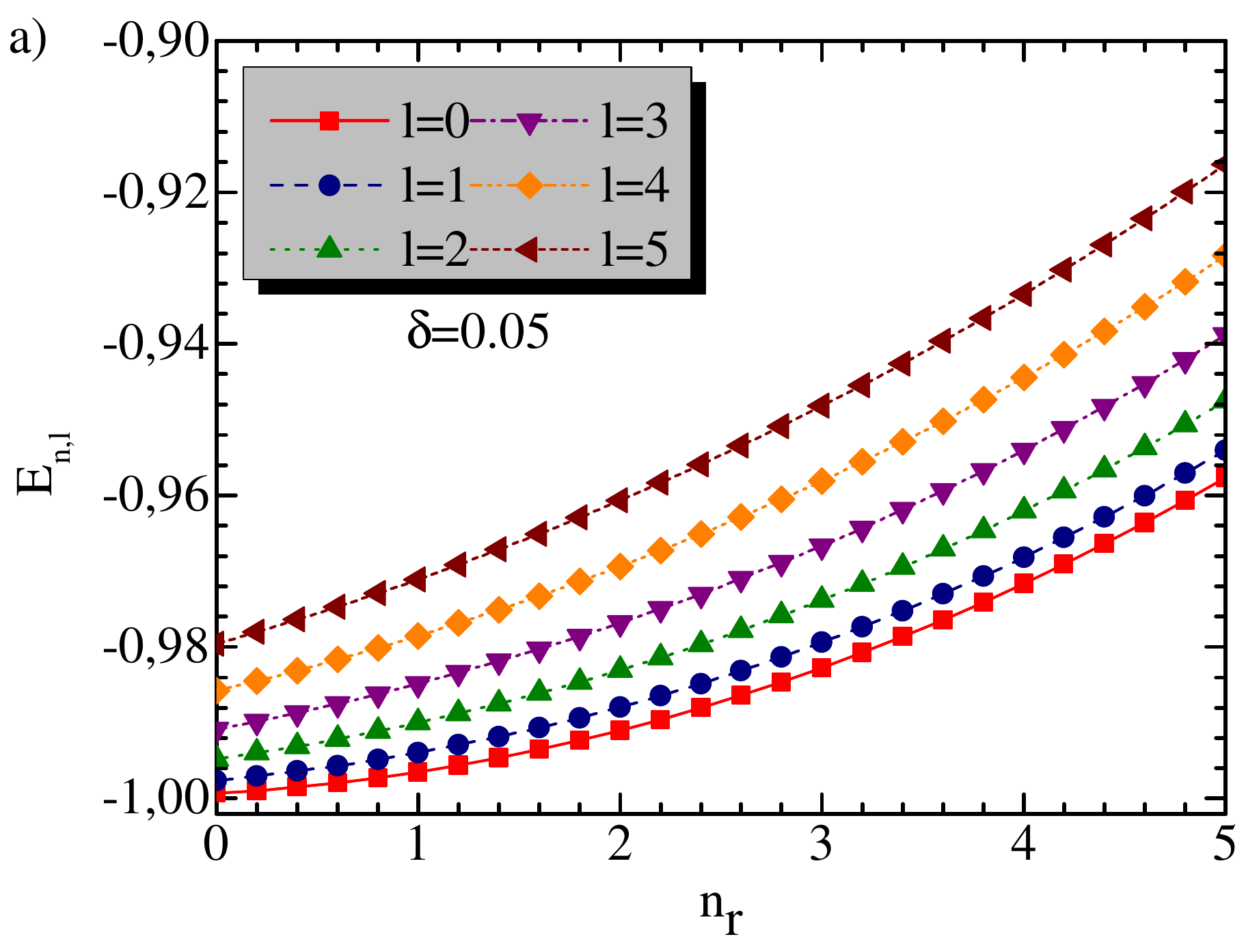}
\includegraphics[scale=0.45]{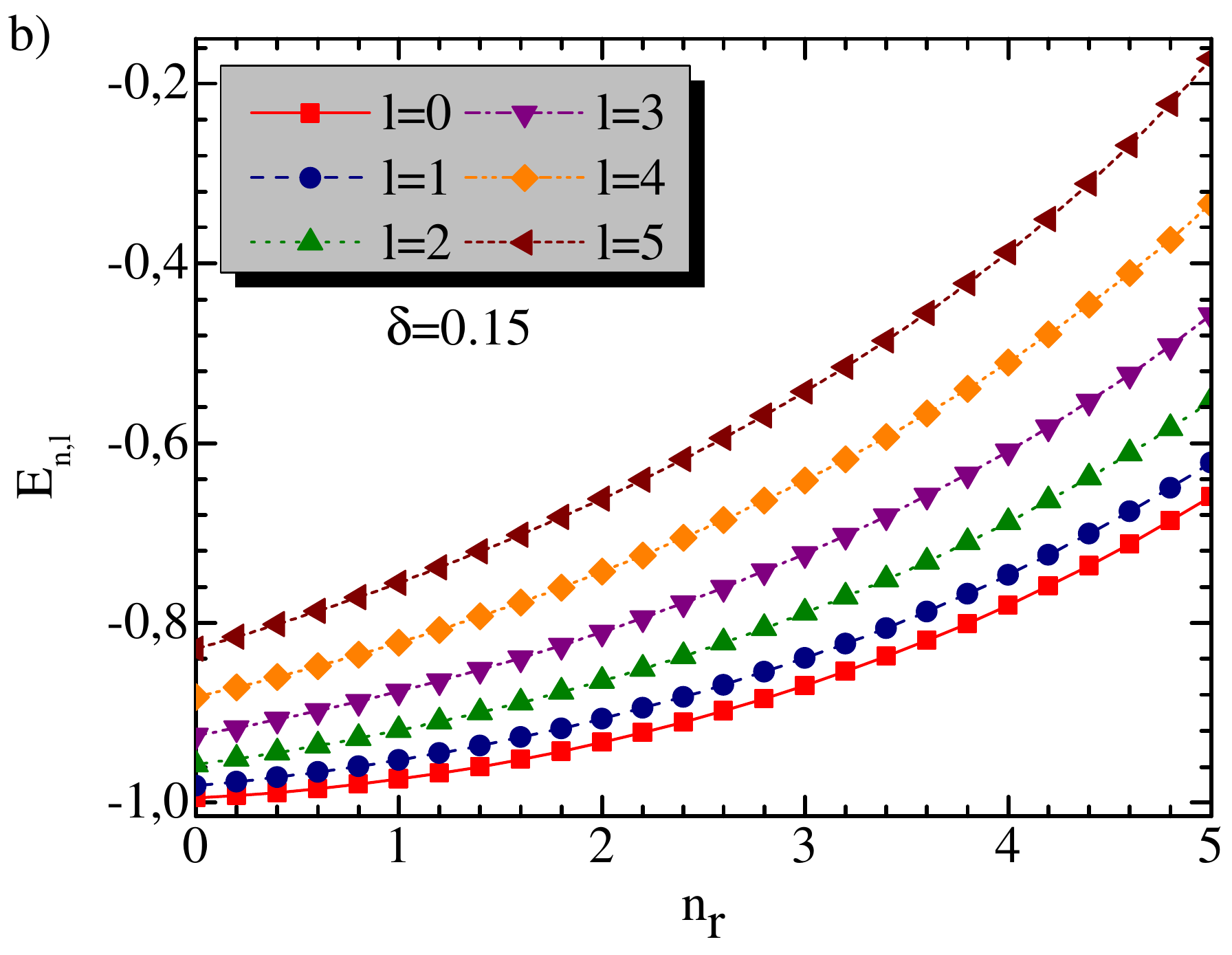}
     \end{center}
          \vspace{-2mm}
\caption{Energy levels $E_{n_r,\ell}$ as a function of $n_r$ with given values of $\ell$ for a) $\delta=0.05$ and b) $\delta=0.15$.\label{fig:Enr}}
\end{figure}

In Fig.~\ref{fig:Enr}, we show the dependence of energy eigenvalues on the quantum number $n_r$ for a given of $\ell=0,..,5$ with $\delta=0.05$ and $0.15$. For all values of $\ell$, the energy eigenvalues increase as the $n_r$ gets bigger. Note that the increment in the case of $\delta=0.15$ is much larger
than in the case of $\delta=0.05$.

\begin{figure}[hbt]
    \begin{center}
\includegraphics[scale=0.45]{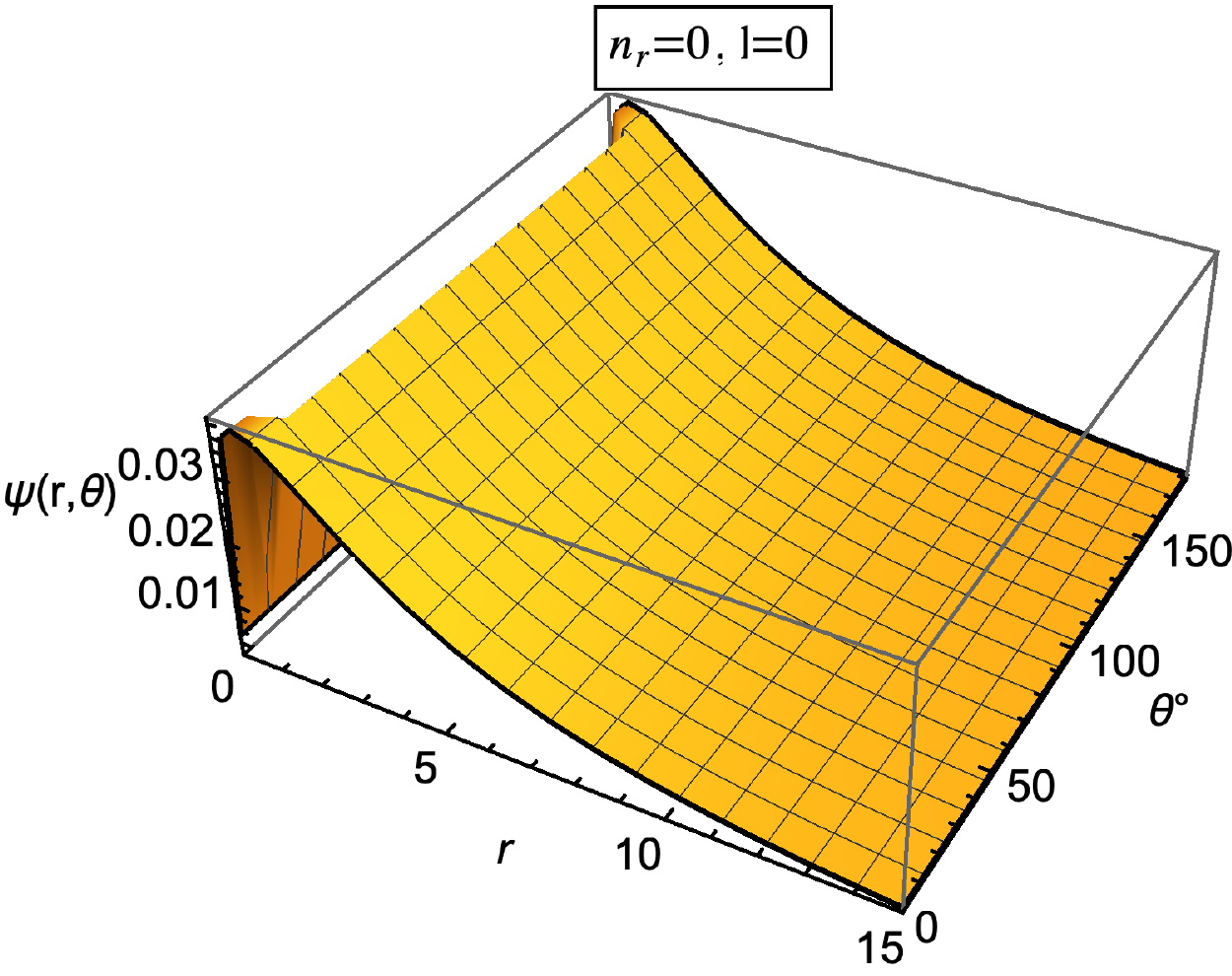}
\includegraphics[scale=0.45]{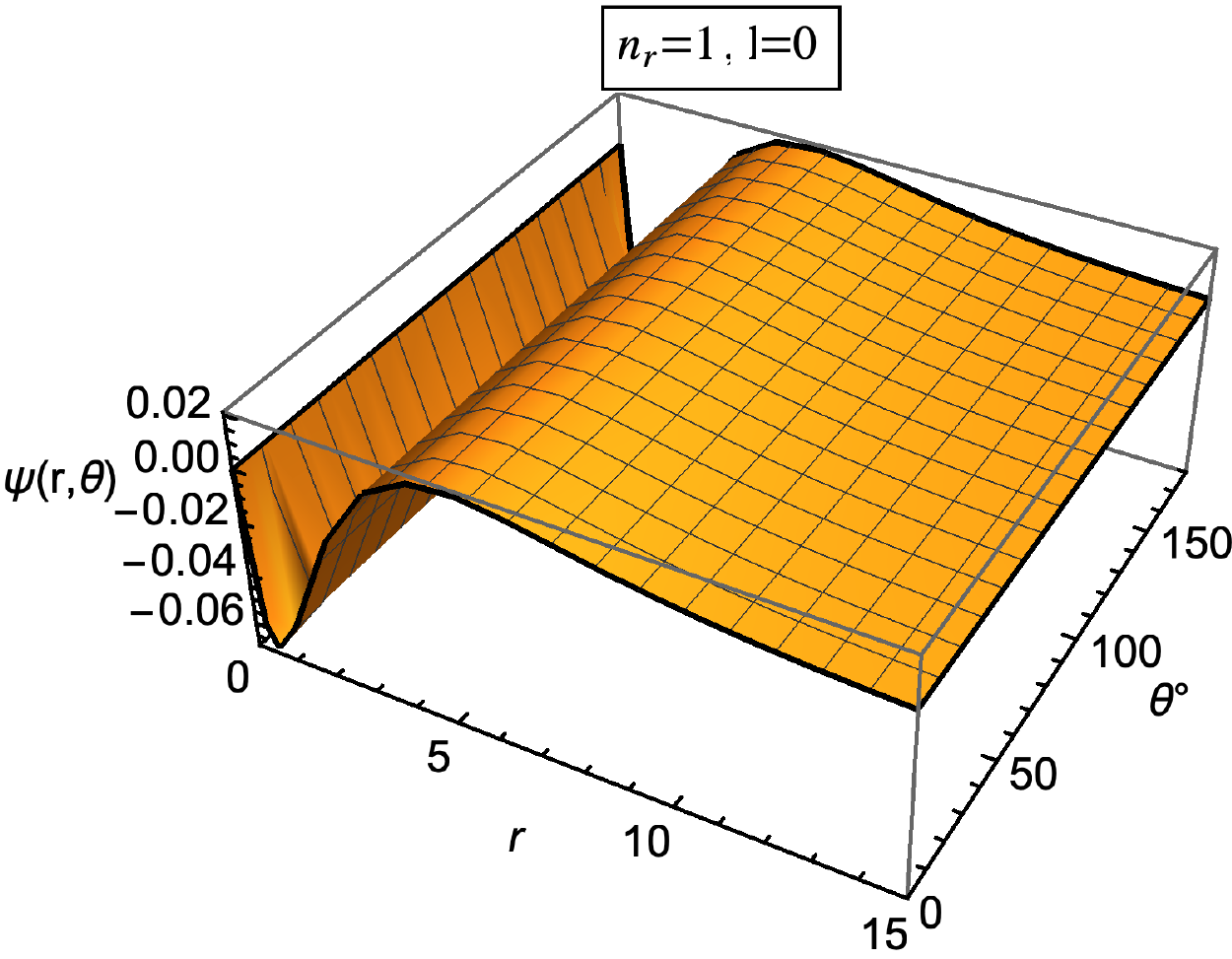}
\includegraphics[scale=0.45]{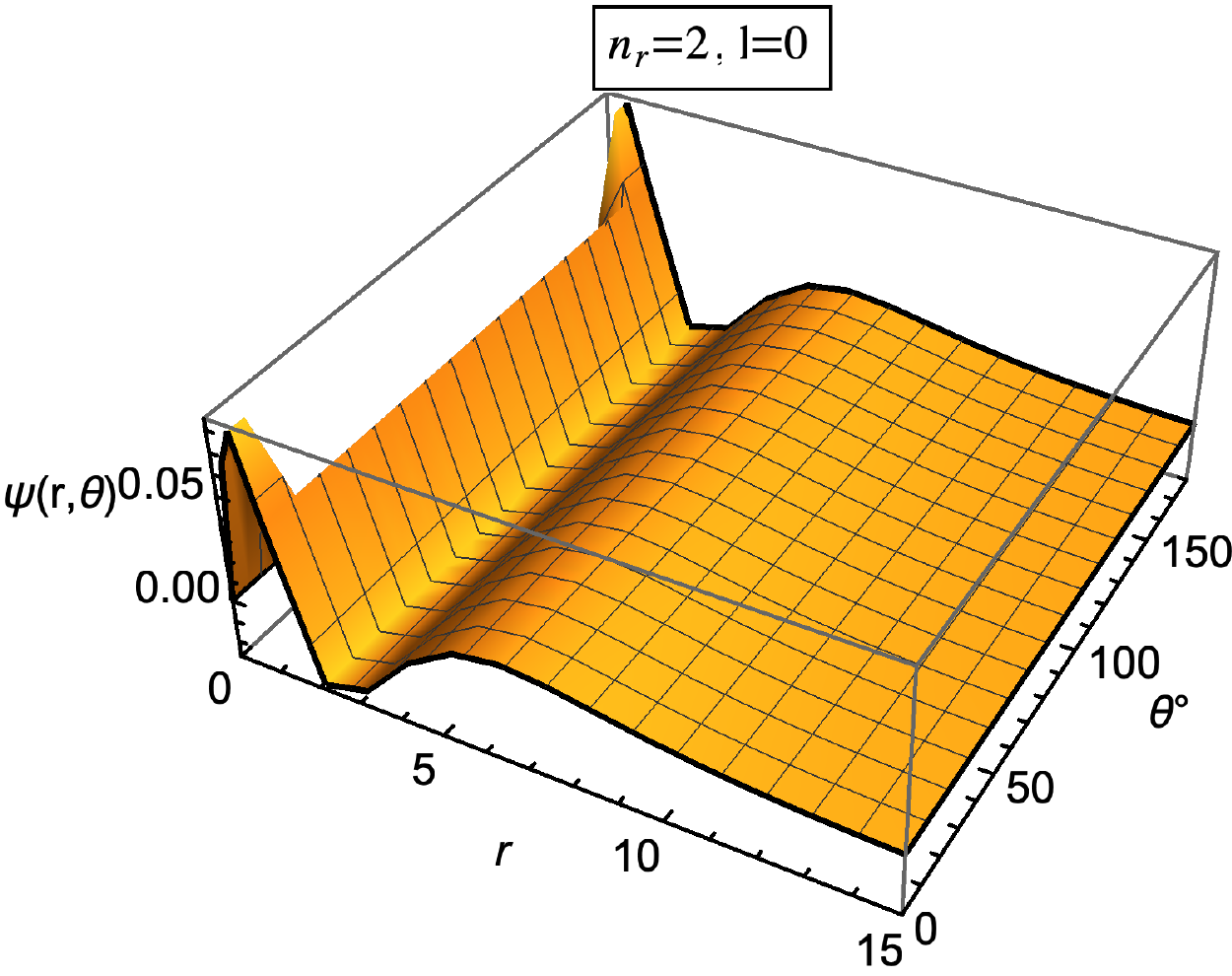}
\includegraphics[scale=0.45]{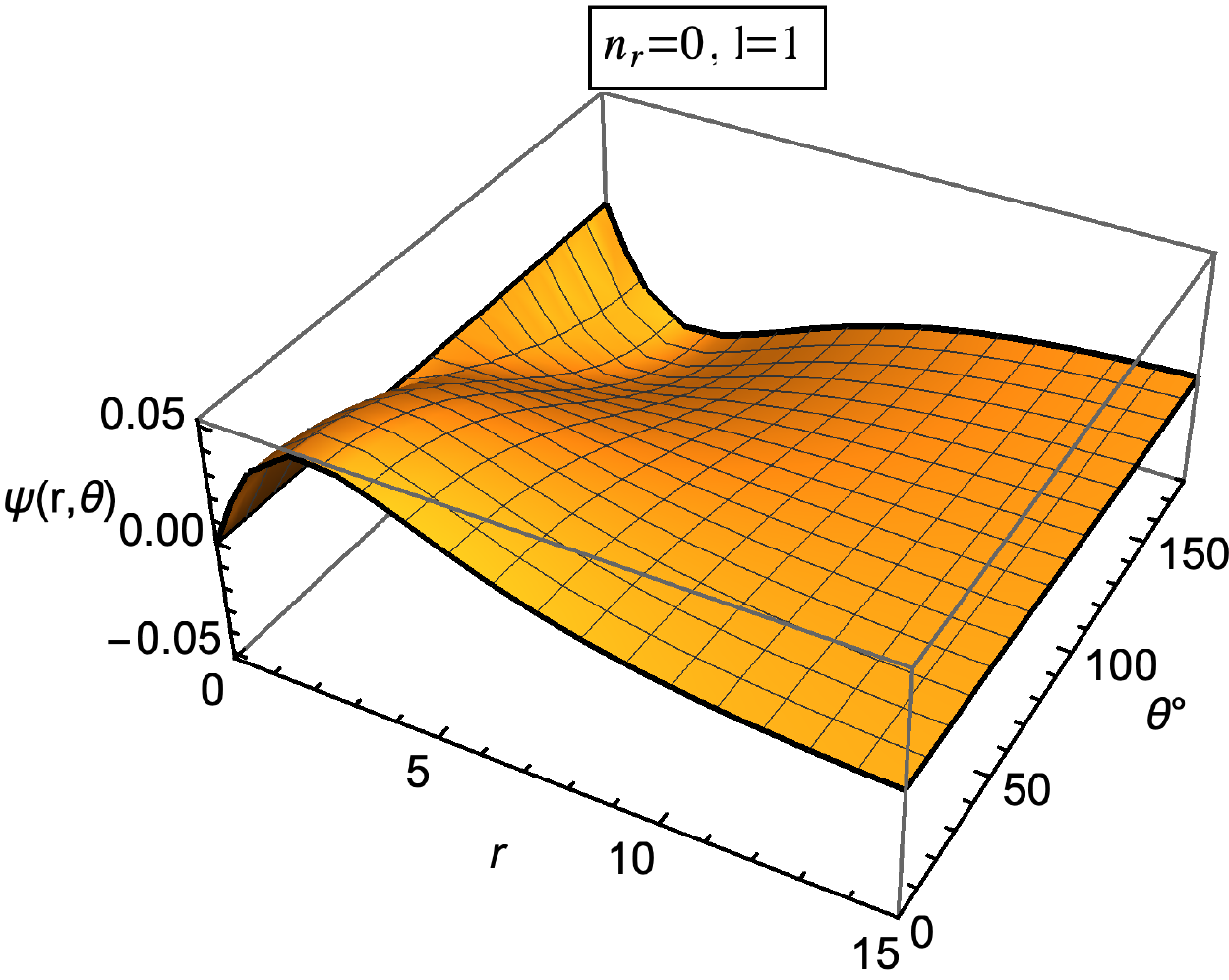}
\includegraphics[scale=0.45]{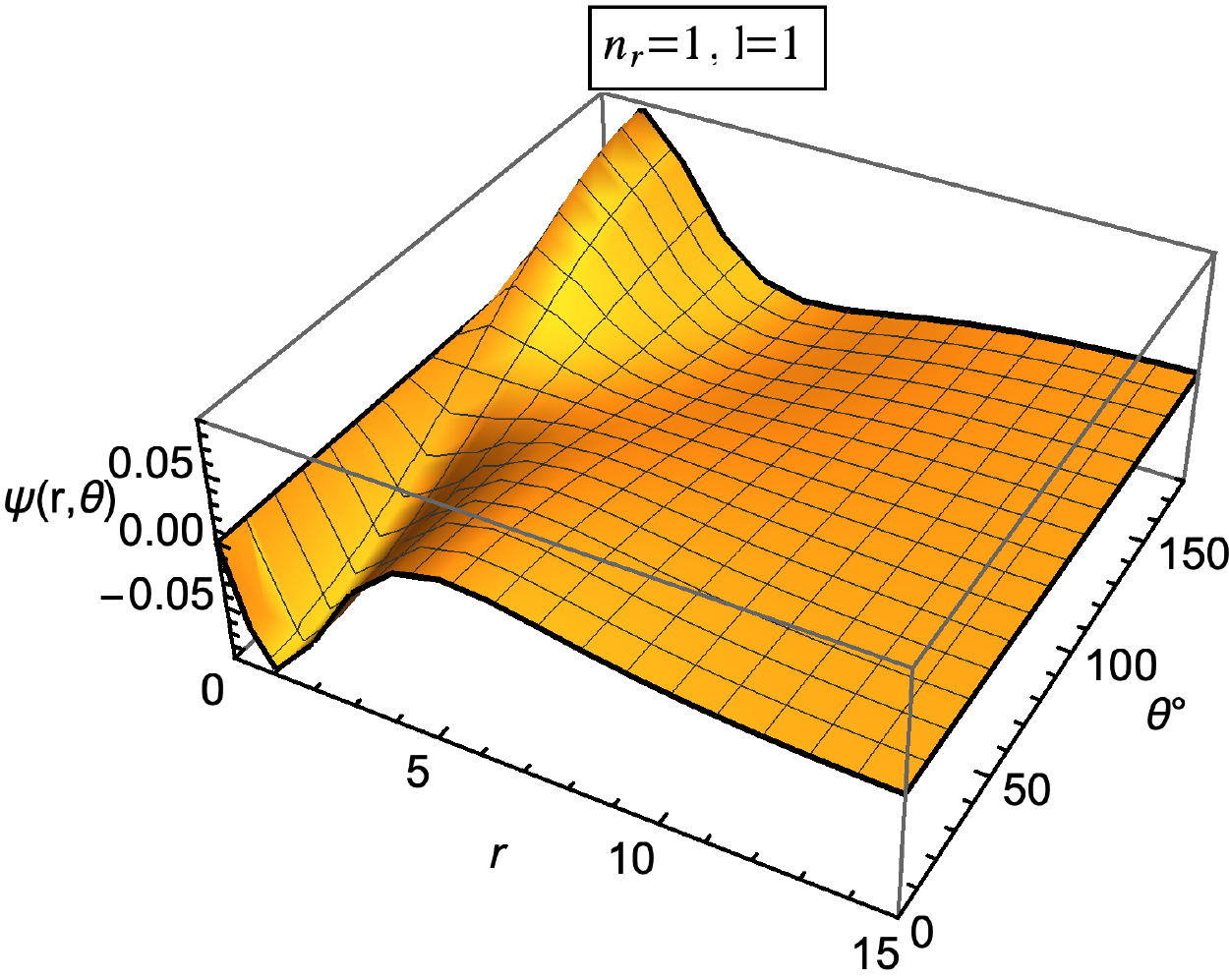}
\includegraphics[scale=0.45]{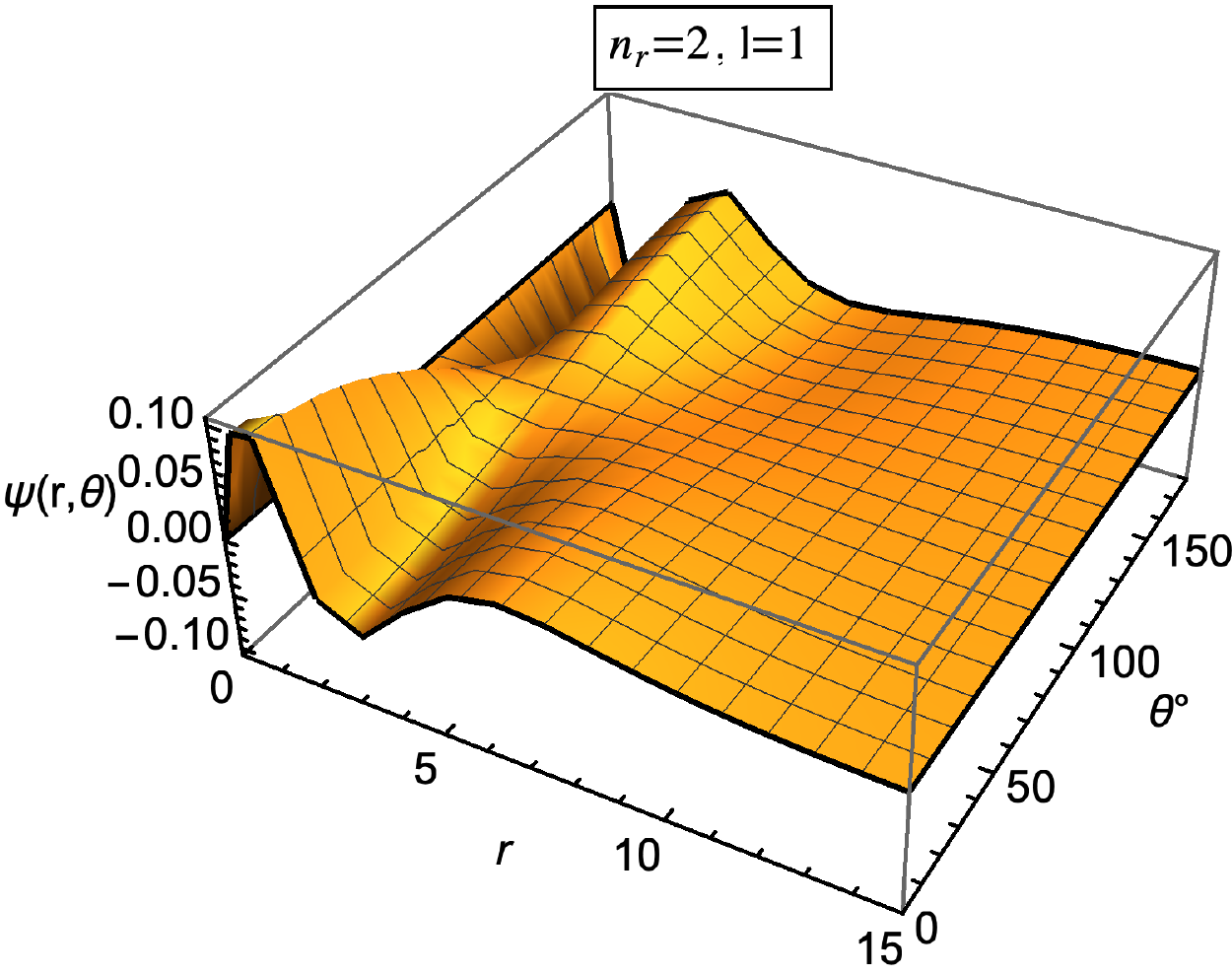}
\includegraphics[scale=0.45]{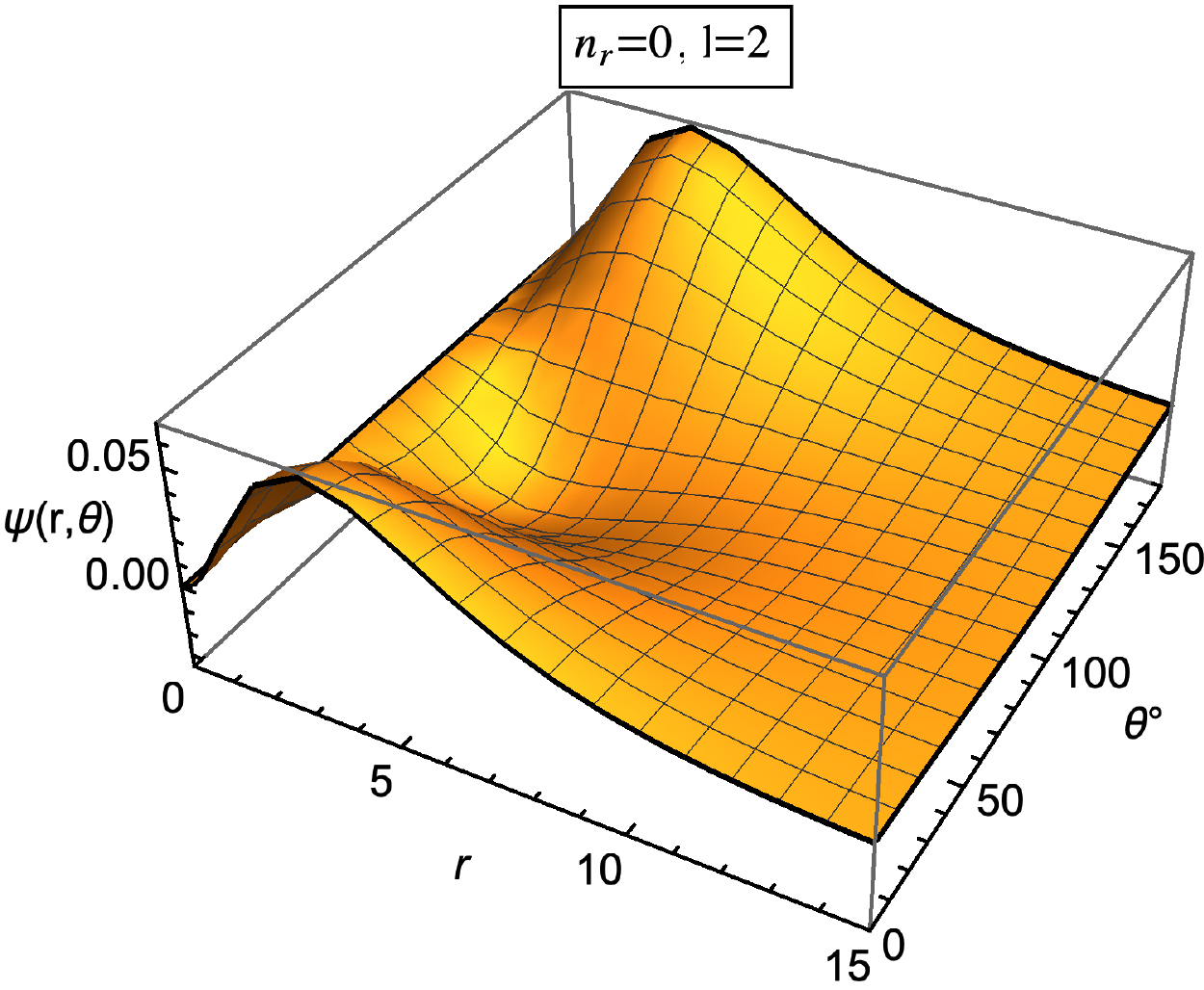}
\includegraphics[scale=0.45]{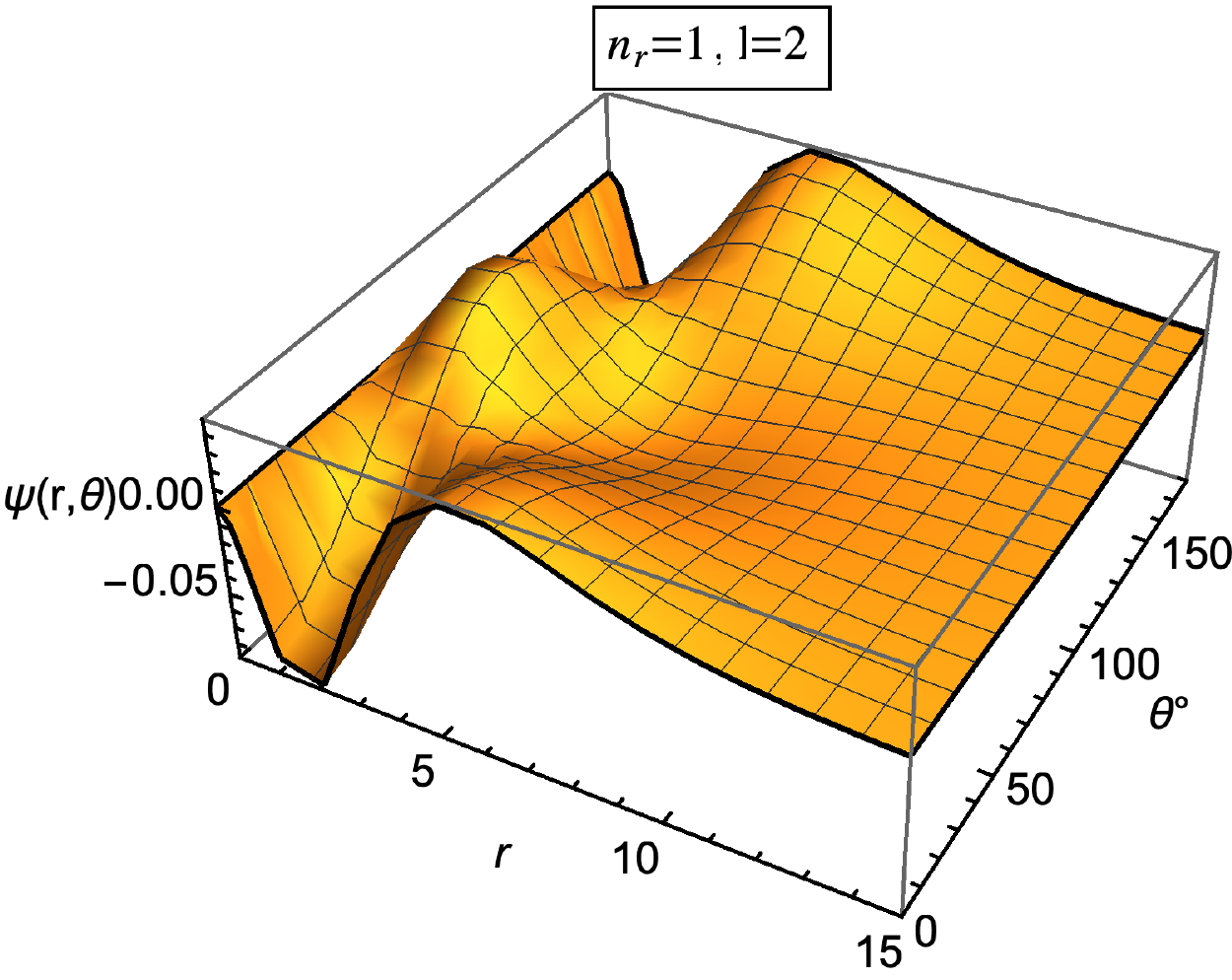}
\includegraphics[scale=0.45]{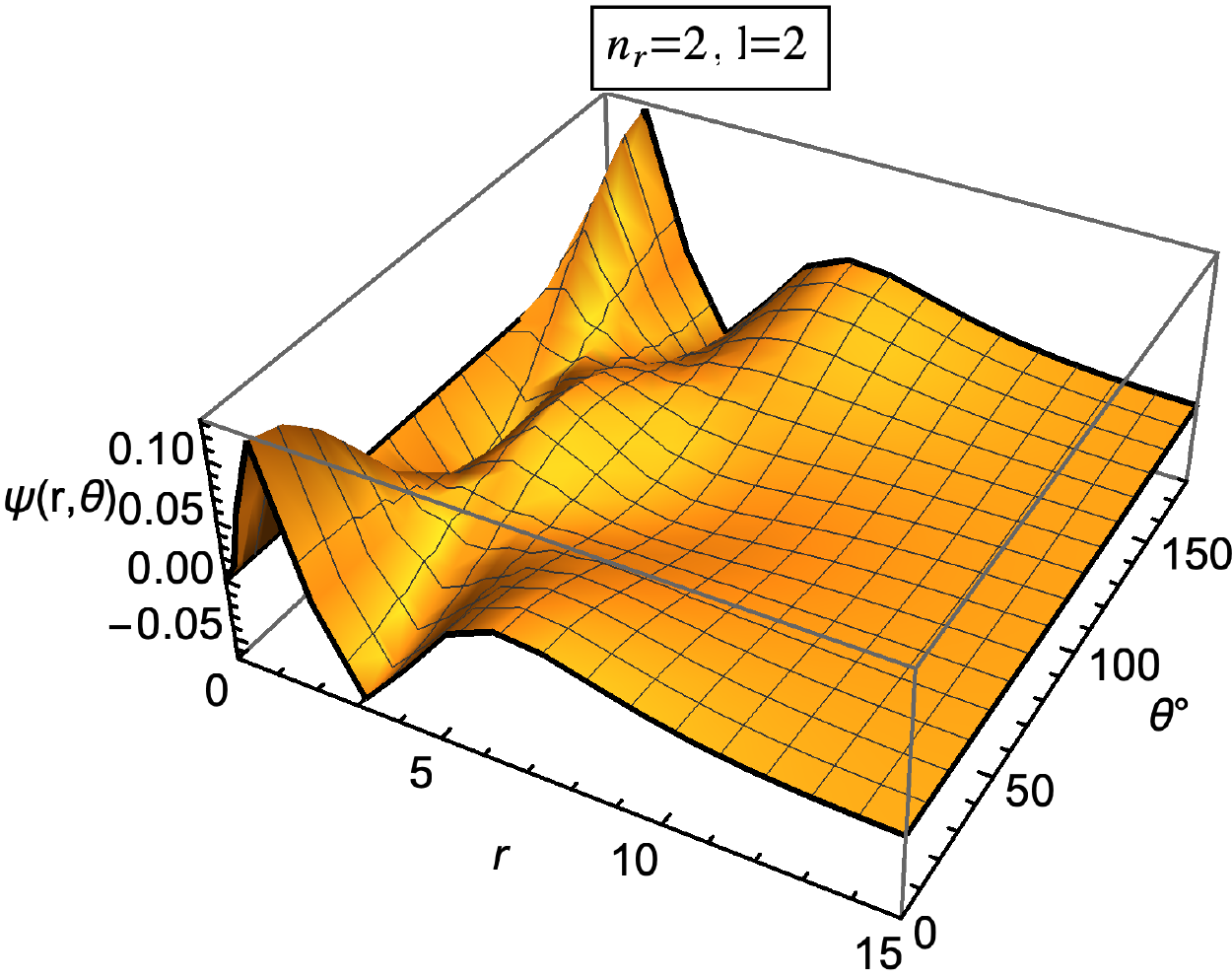}
\includegraphics[scale=0.45]{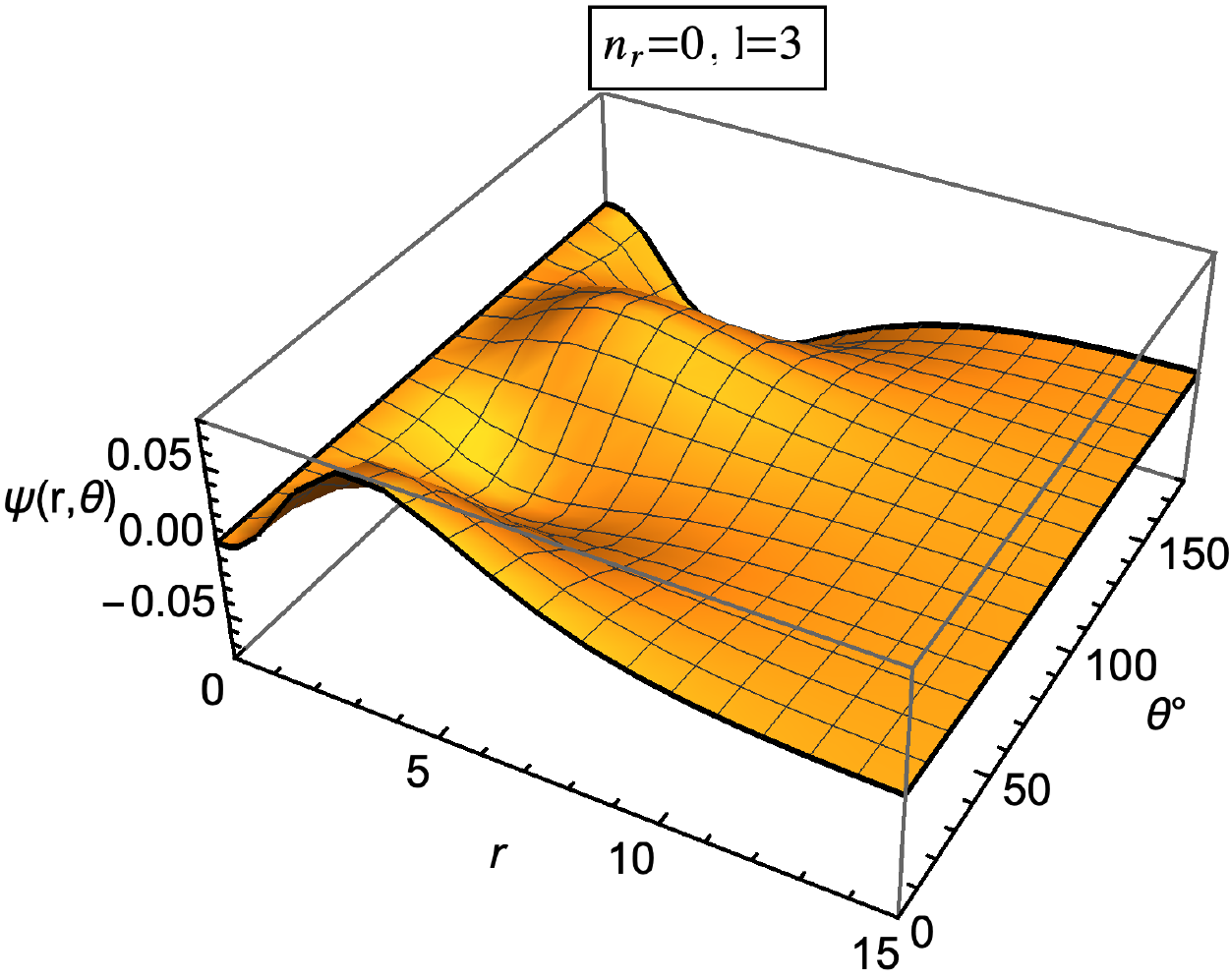}
\includegraphics[scale=0.45]{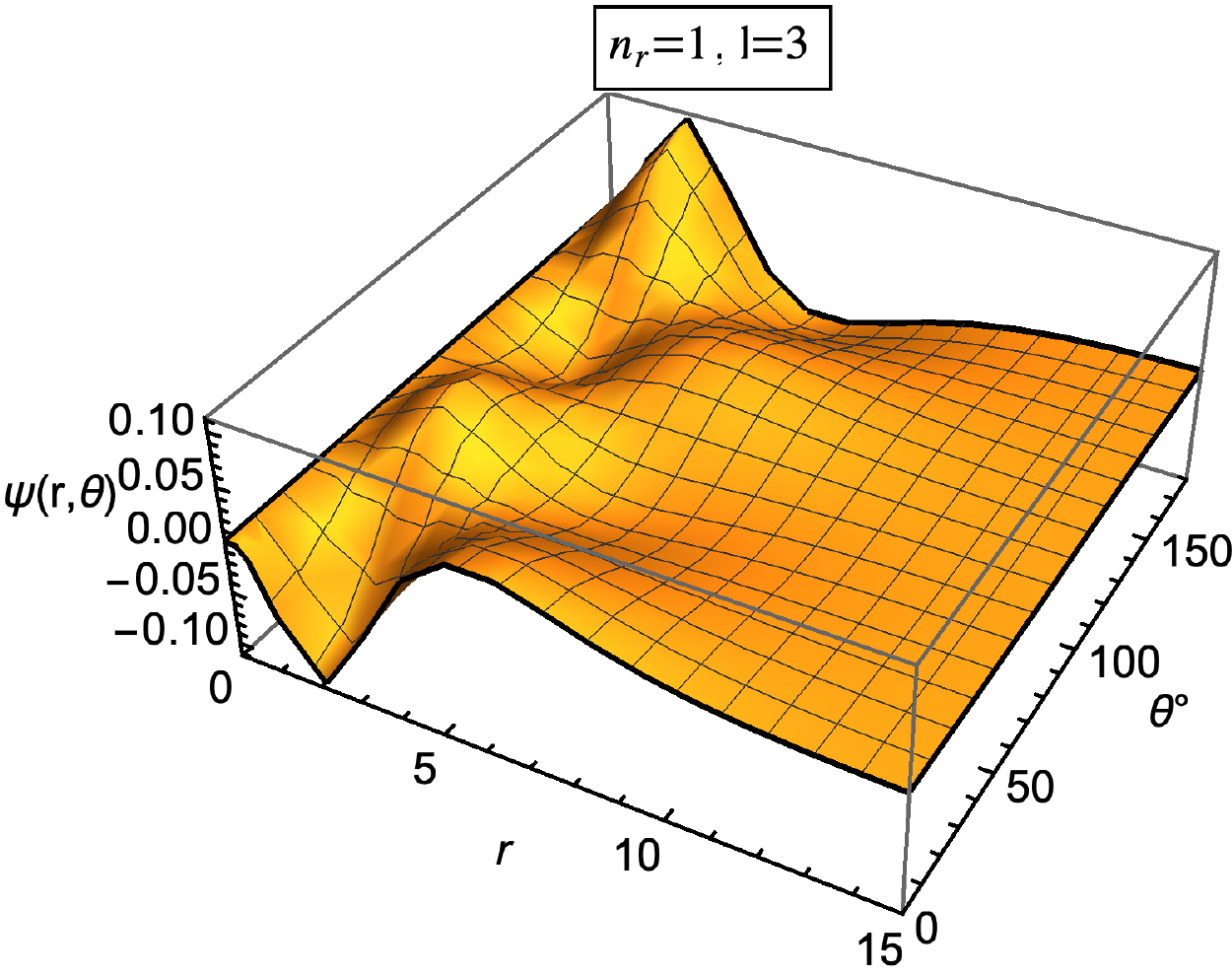}
\includegraphics[scale=0.45]{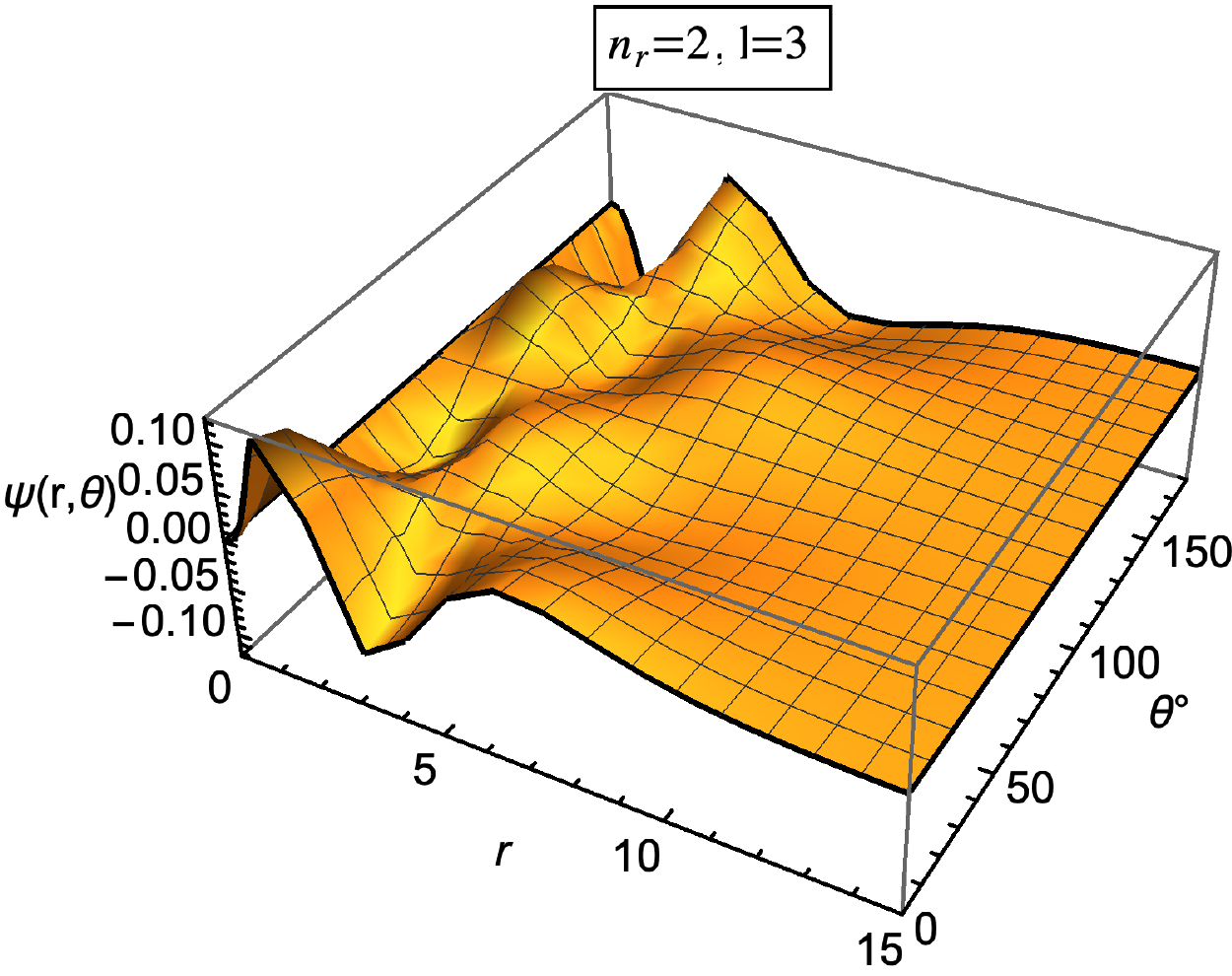}
     \end{center}
 \vspace{-4mm}
\caption{Normalized wave functions as a function of $r$ and $\theta$ for values of $n_r=0,1,2$ and $l=0,1,2,3$.}\label{fig:PsinrN}
\end{figure}

We show in Fig.~\ref{fig:PsinrN} total wave functions $\psi_{n_r,\ell}(r,\theta)$ as a function of $r$ and $\theta$ for $n_r = 0, 1, 2, 3$ and $\ell = 0, 1,2,3$. Here, we vary the $r$ and $\theta$ in the ranges of $r \in [0,15]$ and $\theta \in [0,\pi]$, respectively. We also set the parameter $\delta$ as $\delta=0.15$. 
It is clearly seen from these figures that the wave function has $(n+1)$ and $(\ell+1)$ nodes with associated to the axes $r$ and $\theta$, separately. The number of nodes is not affect by the position dependence of the potential strength, i.e., $V_1$, $V_2$, $V_3$, $V_4$. However, the magnitude and wavelength of the associated wave function are affected.

\begin{table}[h!]
\caption{Bound state energy eigenvalues of $1s, 2s, 2p, 3p, 3d, 4p, 4d$ and $4f$.}\label{table:BS}
\centering
\begin{tabular}{ccccccccc}
\hline
$\delta$ & 1s & 2s & 2p& 3p& 3d& 4p& 4d & 4f\\
\hline\hline
0.05& -0.99649604 & -0.99099814 & -0.98794323 & -0.97940763 & -0.97382843 & -0.96814486& -0.96204162& -0.95403227\\
0.10& -0.98722148 & -0.96718290 & -0.95527801 & -0.92347287 & -0.90142687 & -0.88075546& -0.85604396& -0.82343121\\
0.15& -0.97386879 & -0.93295875 & -0.90640172 & -0.83931973 & -0.78898097 & -0.74667285& -0.68788398& -0.60890610\\
0.20& -0.95804186 & -0.89278636 & -0.84524937 & -0.73363358 & -0.64058725 & -0.57365805& -0.45761787& -0.29378707\\
\hline
\end{tabular}
\end{table}
We present in Table~\ref{table:BS} the energy eigenvalues of $1s, 2s, 2p, 3p, 3d, 4p, 4d, 4f$ for some values of $\delta$. Here, we set the principal quantum number as $n =n_r+\ell +1$. We observe that for a given $n$, the energy eigenvalues increase with increment in $\ell$. This means that the energy eigenvalues are more bounded at smaller values of $\ell$. Also, for a unique quantum state, energy eigenvalues increase as $\delta$ increases.

\begin{figure}[h]
    \begin{center}
    \subfigure[]{\includegraphics[scale=0.54]{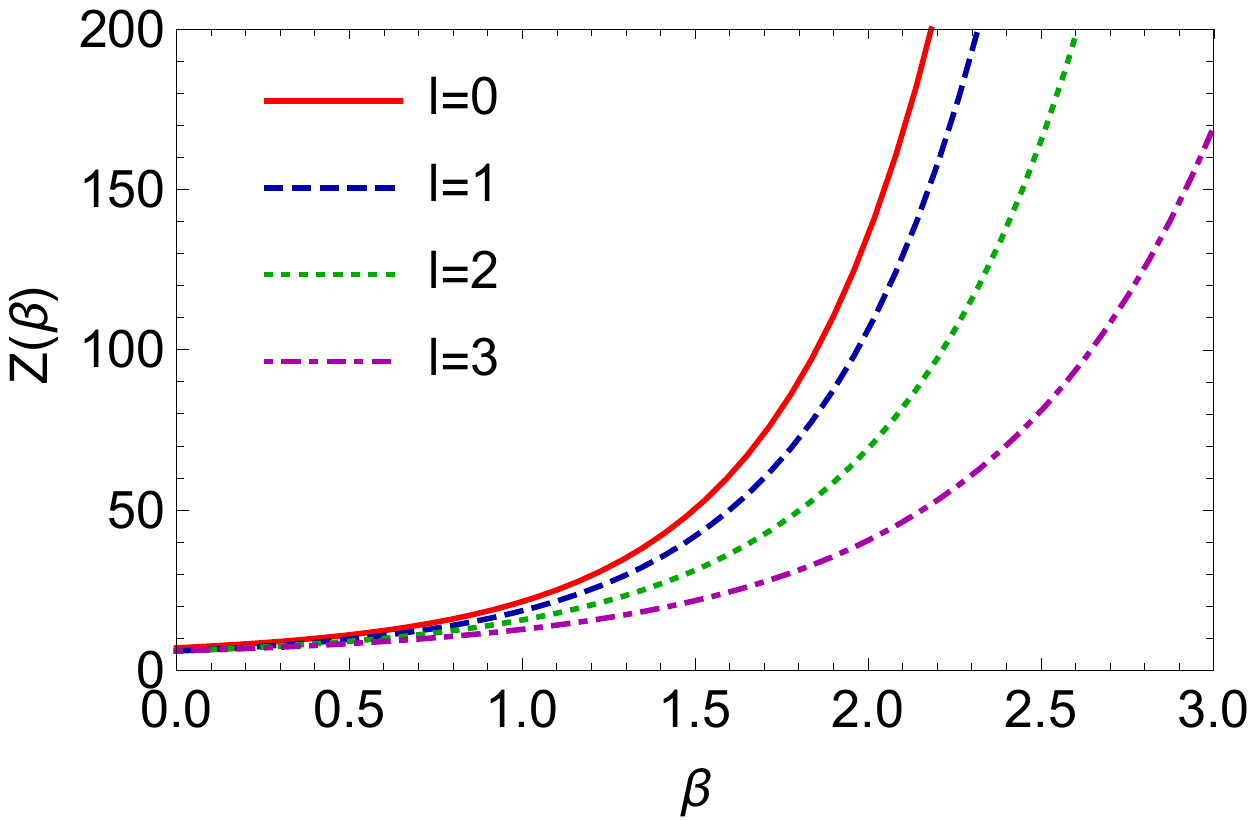}}
	\subfigure[]{\includegraphics[scale=0.54]{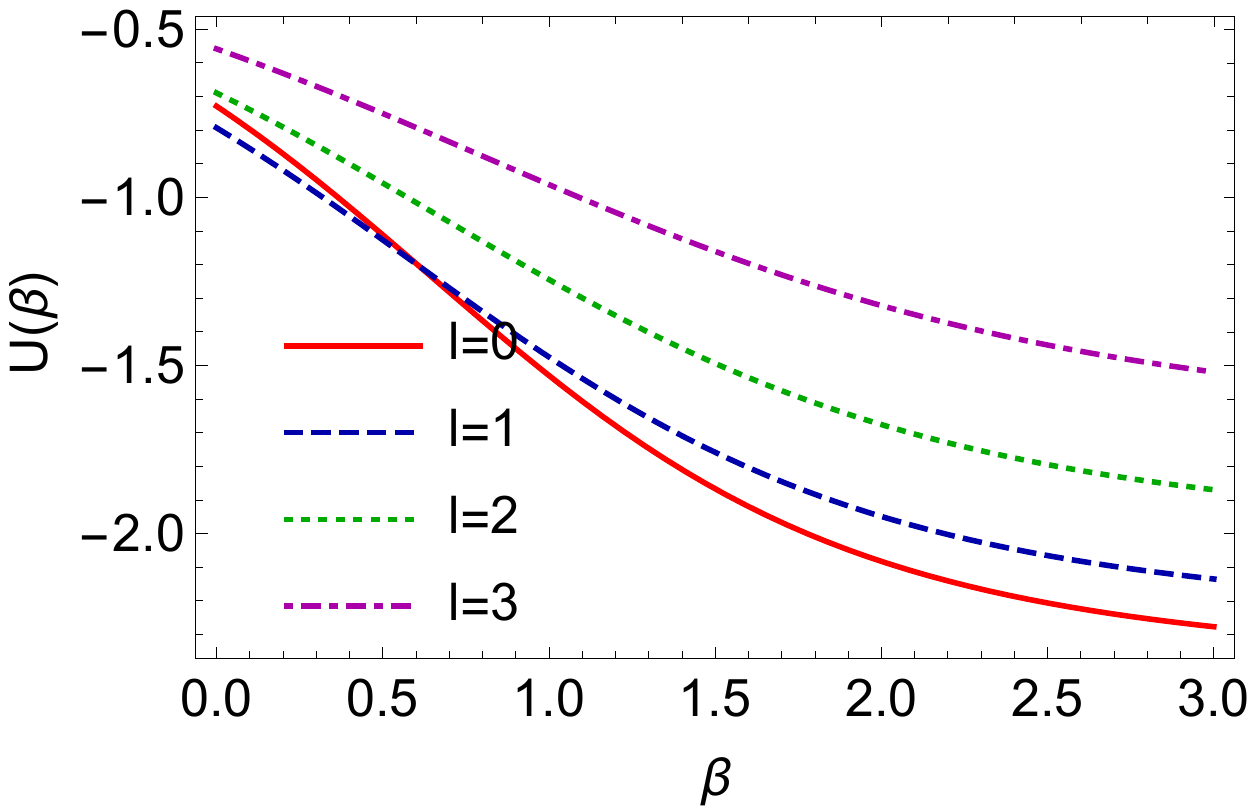}}
	\subfigure[]{\includegraphics[scale=0.54]{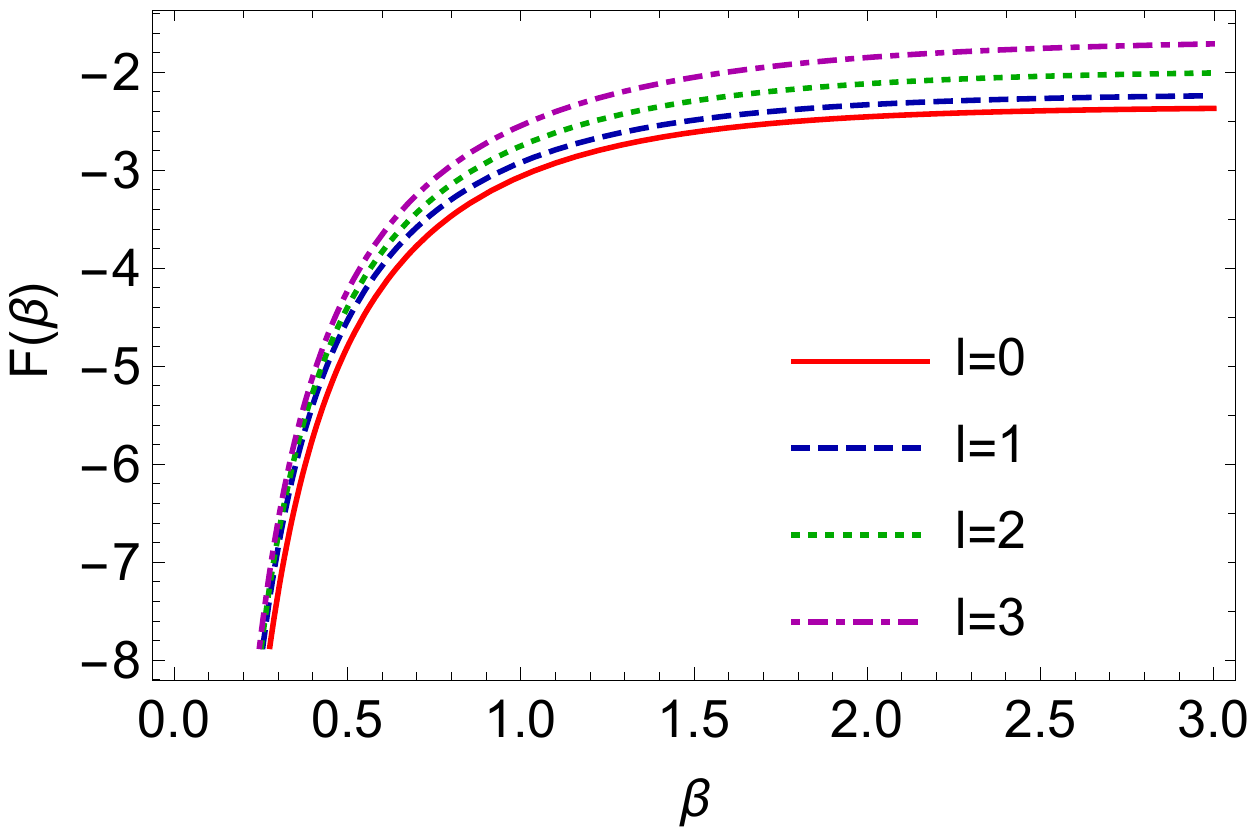}}
	\subfigure[]{\includegraphics[scale=0.54]{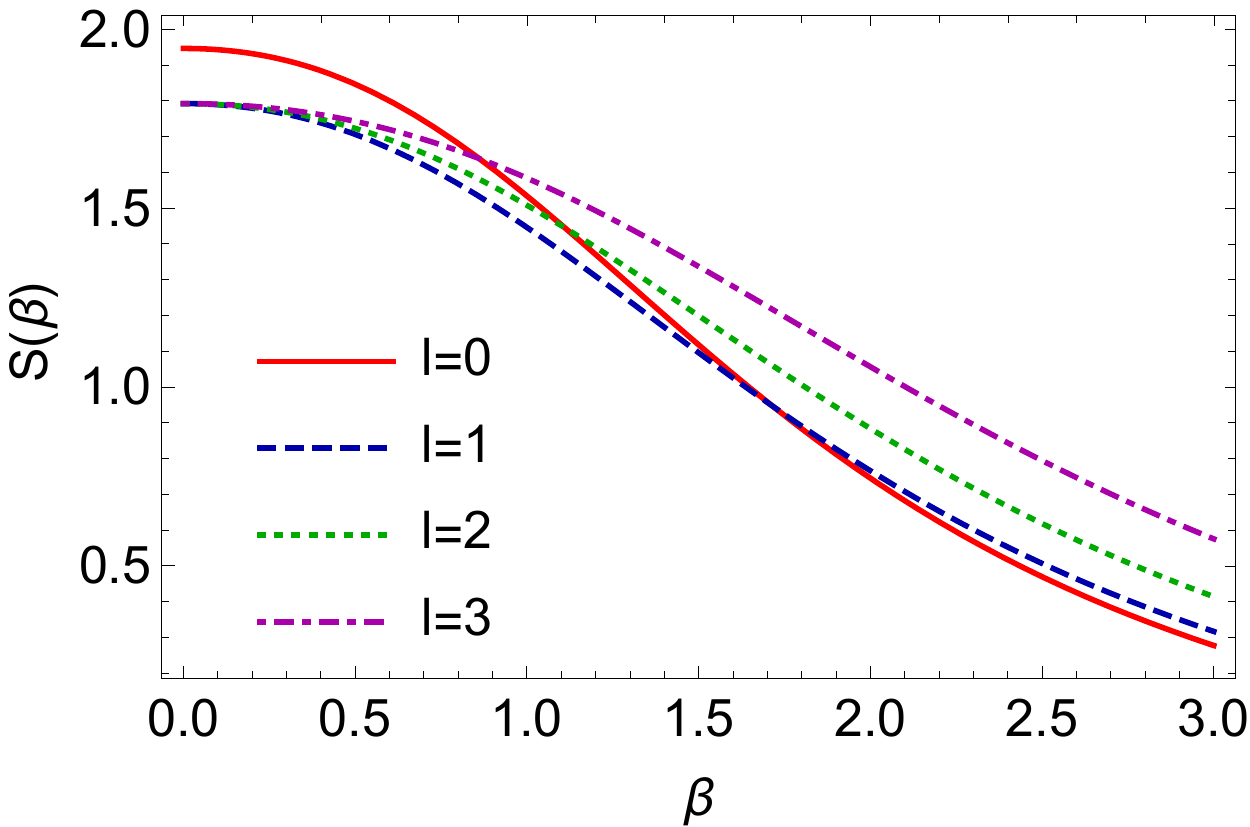}}
	\subfigure[]{\includegraphics[scale=0.54]{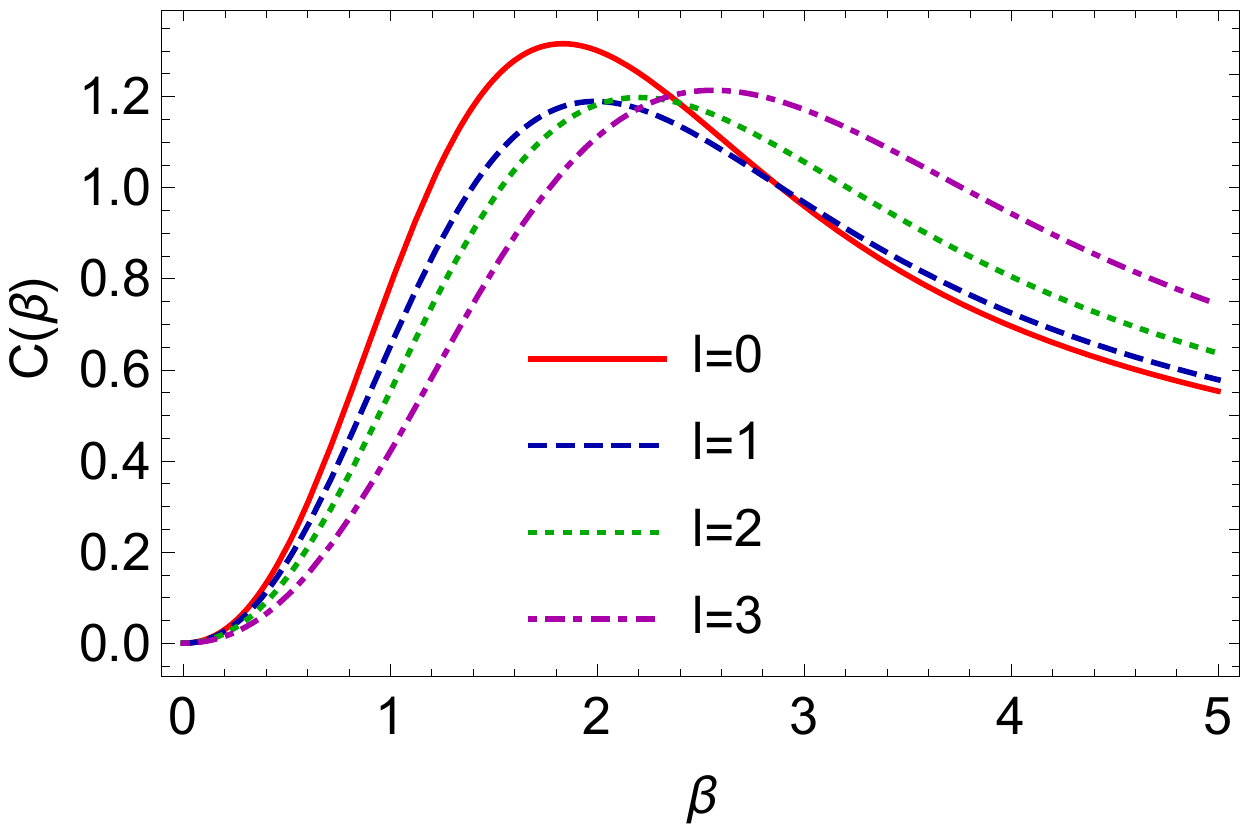}}
     \end{center}
 \vspace{-4mm}
\caption{Thermodynamic quantities as a function of $\beta$ for various $l$ with $\delta=0.15$.}\label{fig:Termall}
\end{figure}

\begin{figure}[ht]
    \begin{center}
    \subfigure[]{\includegraphics[scale=0.54]{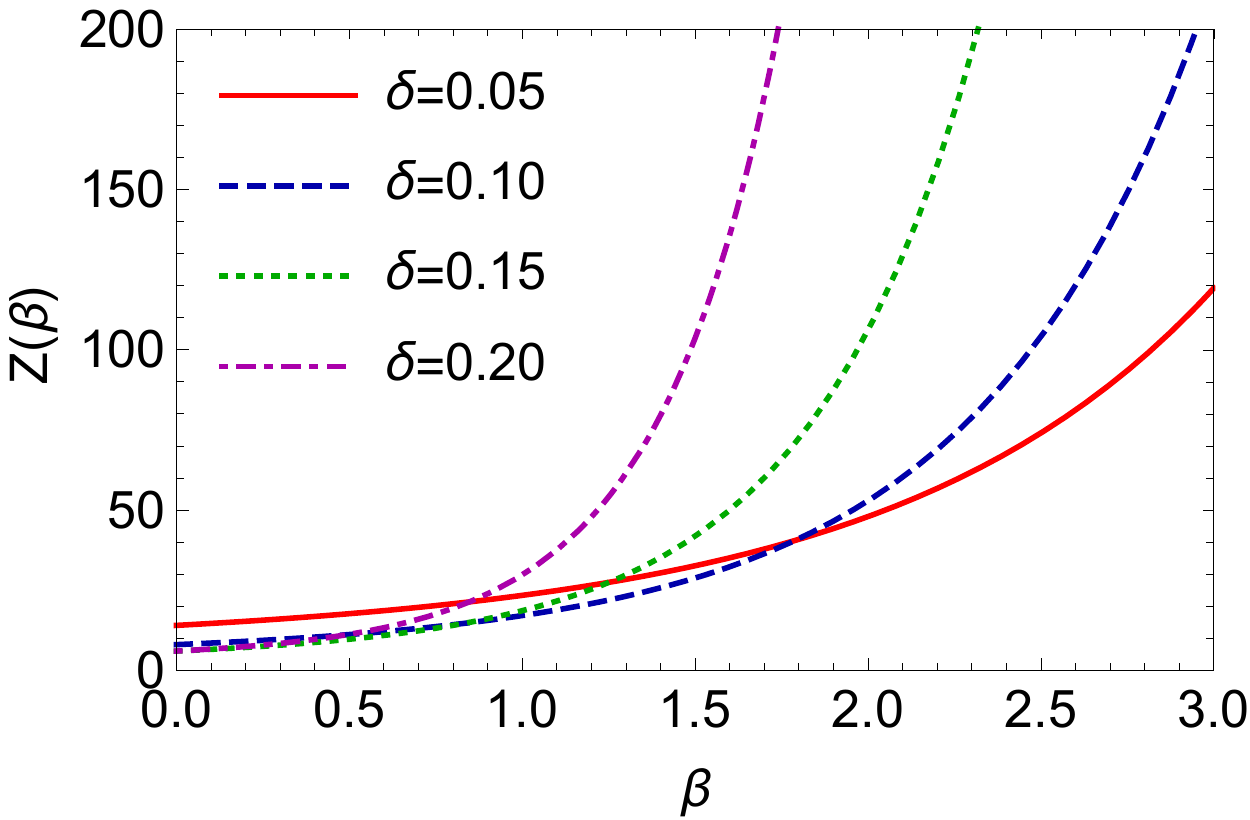}}
	\subfigure[]{\includegraphics[scale=0.54]{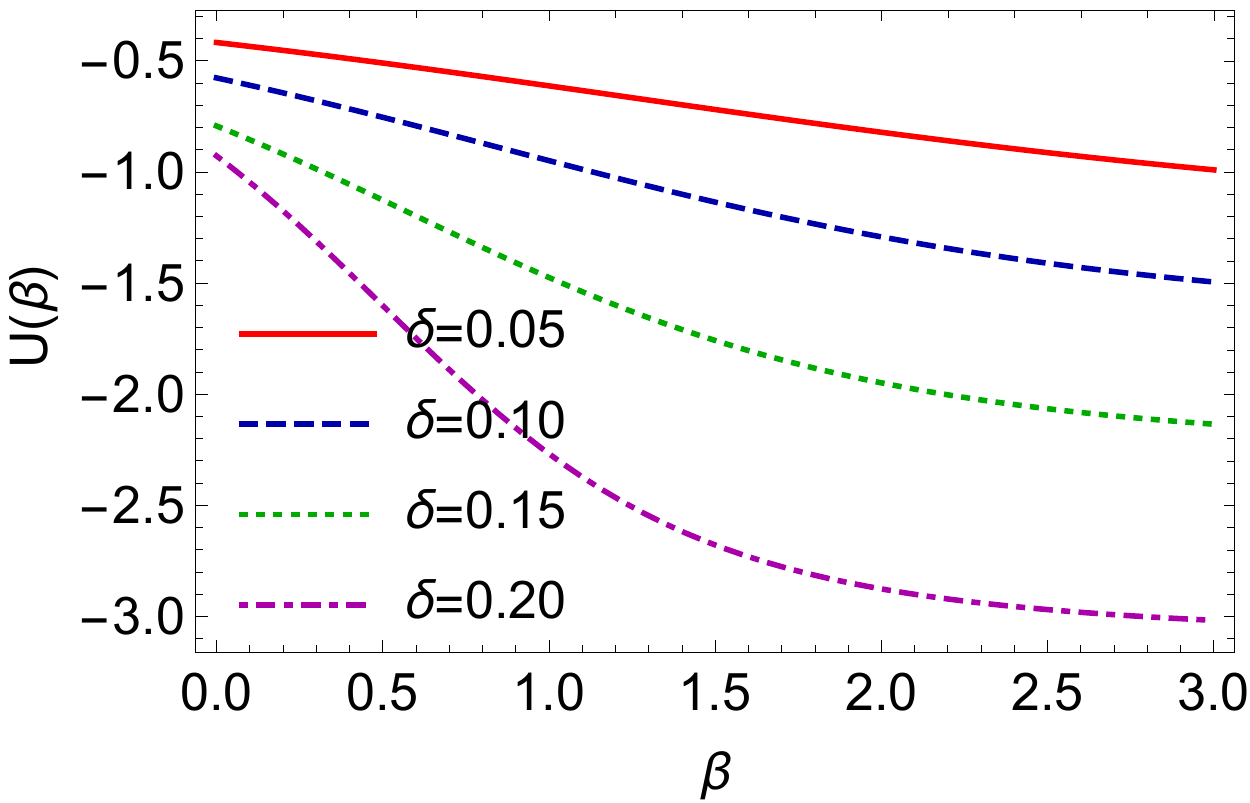}}
	\subfigure[]{\includegraphics[scale=0.54]{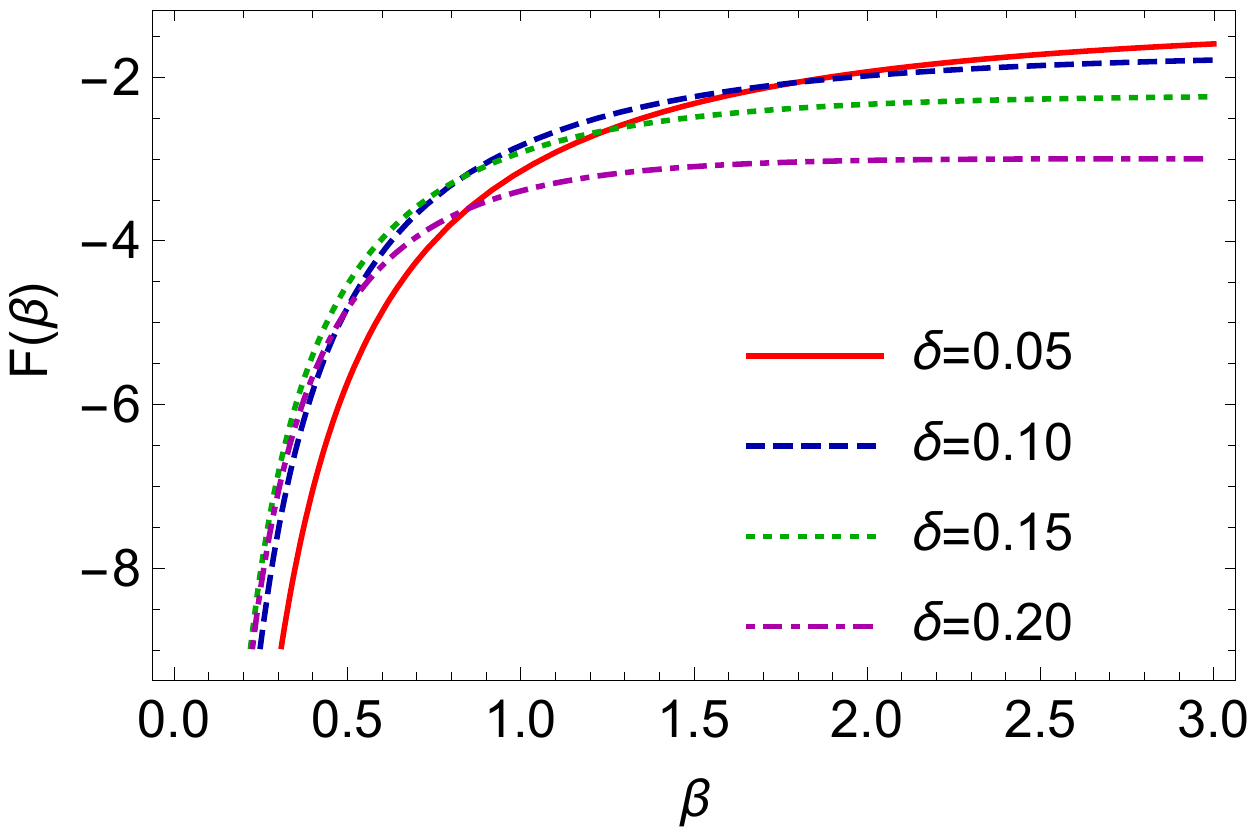}}
	\subfigure[]{\includegraphics[scale=0.54]{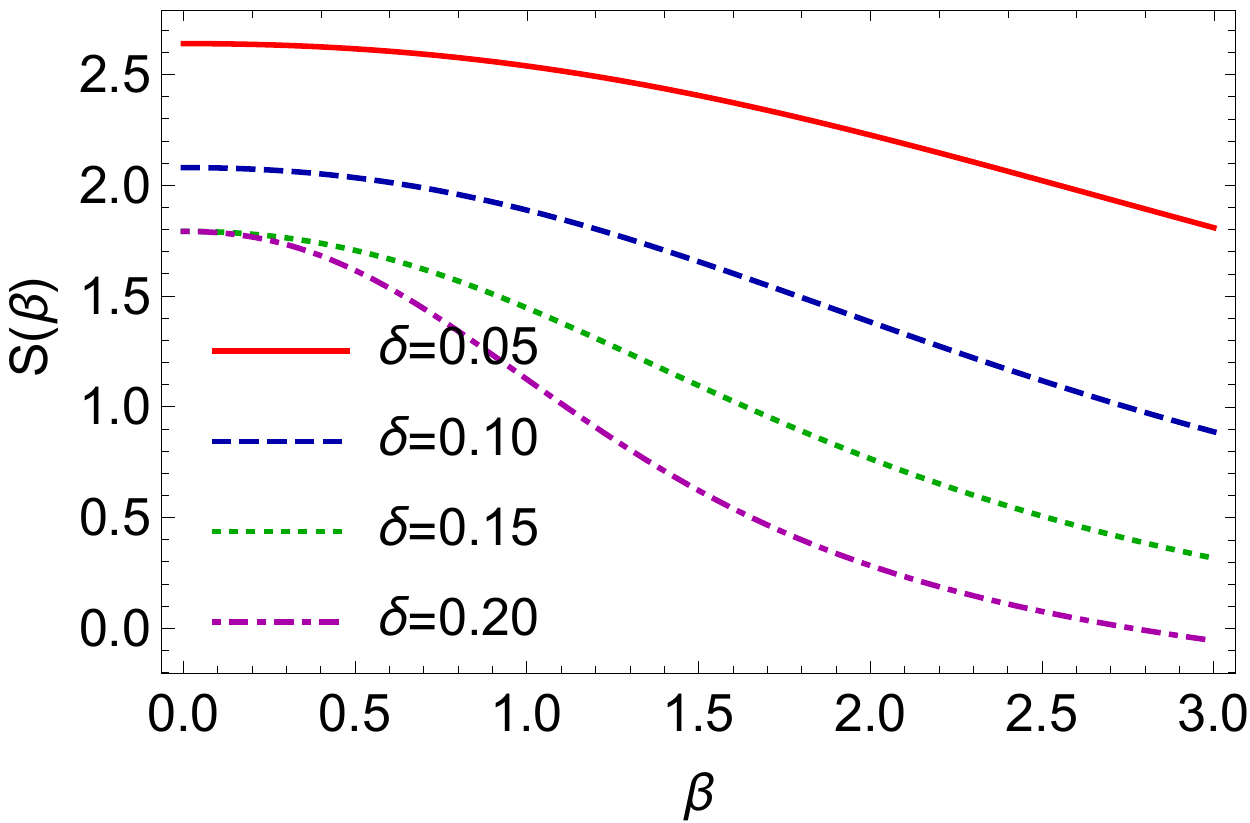}}
	\subfigure[]{\includegraphics[scale=0.54]{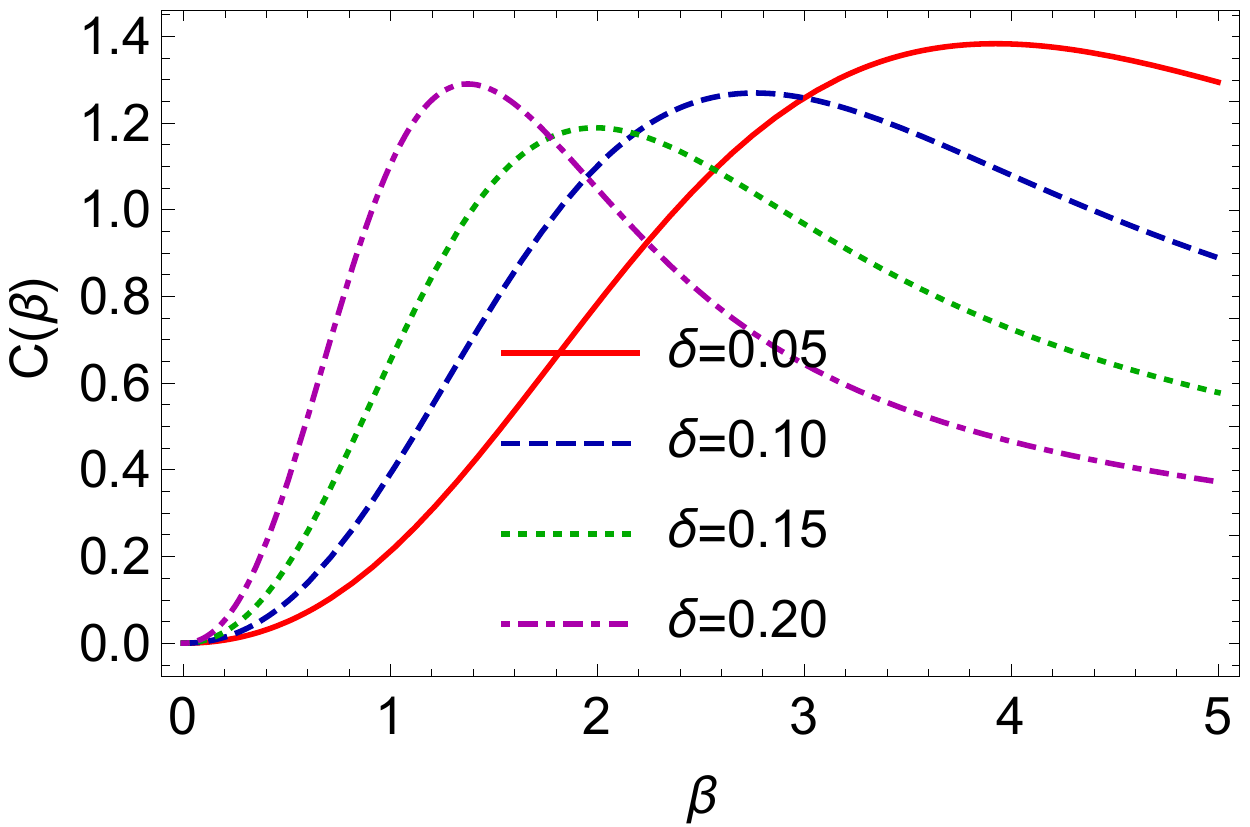}}
     \end{center}
 \vspace{-4mm}
\caption{Thermodynamic quantities as a function of $\beta$ for various $\delta$ and $l=1$.}\label{fig:Termal_del}
\end{figure}

Second, we analyze the dependence of thermodynamic quantities at non-relativistic limit, including partition function $Z(\beta)$, mean energy $U(\beta)$, free energy $F(\beta)$, entropy $S(\beta)$, and specific heat capacity $C(\beta)$, on the parameter $\beta=\frac{1}{k_BT}$. We plot these quantities in Fig.~\ref{fig:Termall} and Fig.~\ref{fig:Termal_del} for different values of $\ell$ and $\delta$, respectively. 
In Fig.~\ref{fig:Termall}, we determine the upper bound vibration quantum number from Eq.\eqref{Eq:lambda} as $N=6$ for $\ell=0$ and $N=5$ for $\ell=1,2,3$. As mentioned earlier, $N$ is equal to $\lambda$ rounded to an integer. Moreover, in Fig.~\ref{fig:Termal_del}, we obtain the $N$ from Eq.\eqref{Eq:lambda} as $N=13,7,5,5$ for $\delta=0.05,0.10,0.15,0.20$, respectively.

It can be seen from these figures that the partition function $Z(\beta)$ increases exponentially as the $\beta$, i.e., the inverse temperature, increases for each value of $\ell$ and $\delta$. The $Z(\beta)$ becomes larger for either smaller $\ell$ or larger $\delta$. The mean energy $U(\beta)$ decreases almost linearly while free energy $F(\beta)$ increases as the $\beta$ increases. The free energy $F(\beta)$ increases rapidly for an interval of $\beta \in [0,1.0]$ and then becomes a nearly constant value with increment of $\beta$. The entropy $S(\beta)$ stays almost unchanged for an interval of $\beta \in [0,0.5]$ and then continues to decrease rapidly with increment in $\beta$.  The functions of $U(\beta)$, $F(\beta)$ and $S(\beta)$ become usually larger for either larger $\ell$ or smaller $\delta$.  The specific heat capacity $C(\beta)$ increases until a maximum value and then decreases as the $\beta$ increases. Its peaks move to the increment direction of $\beta$ with increasing $\ell$ values or decreasing $\delta$ values. The $C(\beta)$ have larger values up to its peak for either smaller $\ell$ or larger $\delta$, and after its peak, this case is vice versa. 

By using the different potentials to represent the internal vibrations of diatomic molecules, some authors have successfully predicted the thermal properties for real substances \cite{Jia17,Strekalov7,Song17,Jia17b,Ikot23}. Accordingly, we also examine the thermal properties for a few diatomic molecules $LiH$, $HCl$, $CuLi$ and $NiC$ through our potential model. 
\begin{table}[h!]
\caption{Spectroscopic constants of the some diatomic molecules.}\label{table:diatomic}
\centering
\begin{tabular}{ccc}
\hline
Molecules & $\delta (\AA^{-1}) $ & $\mu (amu)$ \\
\hline\hline
$LiH$ & 1.1280 & 0.8801221\\
$HCl$& 1.8677 & 0.9801045\\
$CuLi$& 1.00818 & 6.259494\\
$NiC$& 2.25297 & 9.974265\\
\hline
\end{tabular}
\end{table}
We take the spectroscopic parameters from Refs. \cite{Wen1,Oyewumi13} for these molecules as given in Table~\ref{table:diatomic}. Here, we use the conversion factors as follows: $\hbar c = 1973.296$ eV$\AA$ and $1$ $amu = 931.494028$ MeV/c$^2$.

\begin{figure}[ht]
    \begin{center}
    \subfigure[]{\includegraphics[scale=0.54]{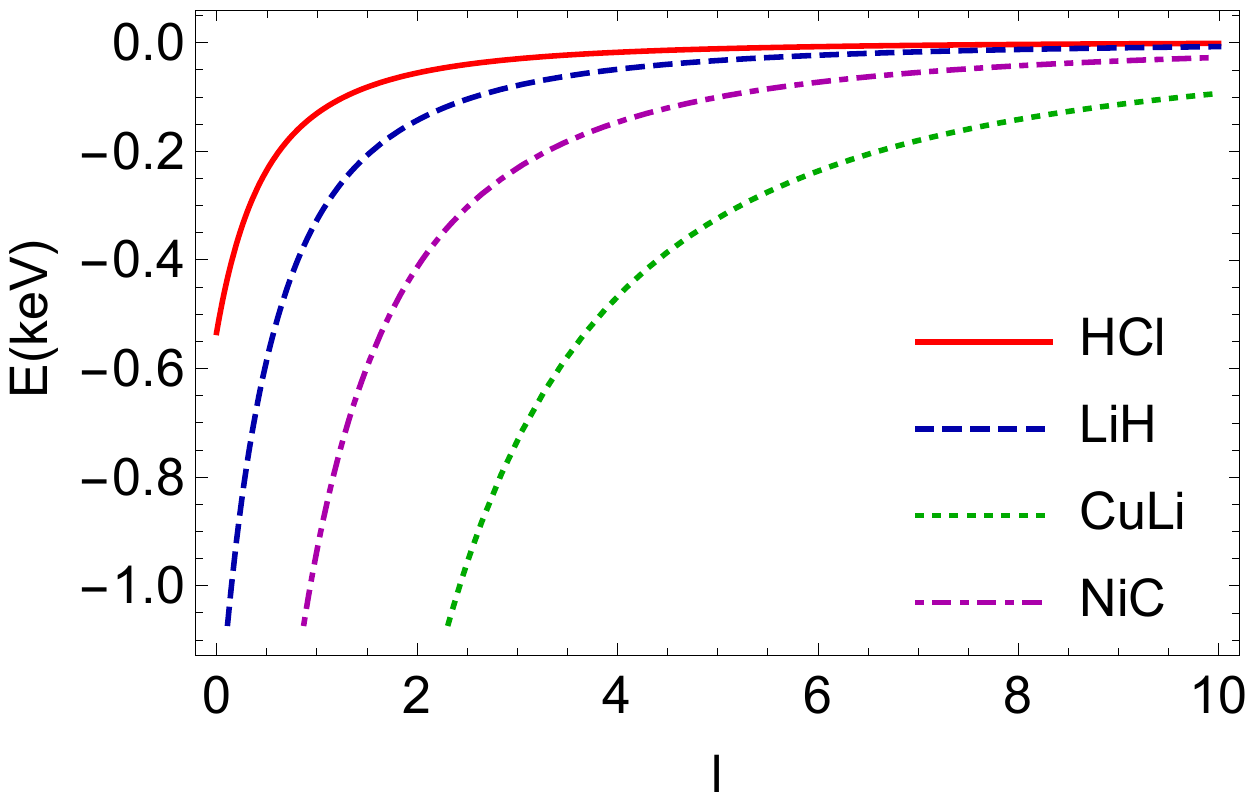}}
    \subfigure[]{\includegraphics[scale=0.54]{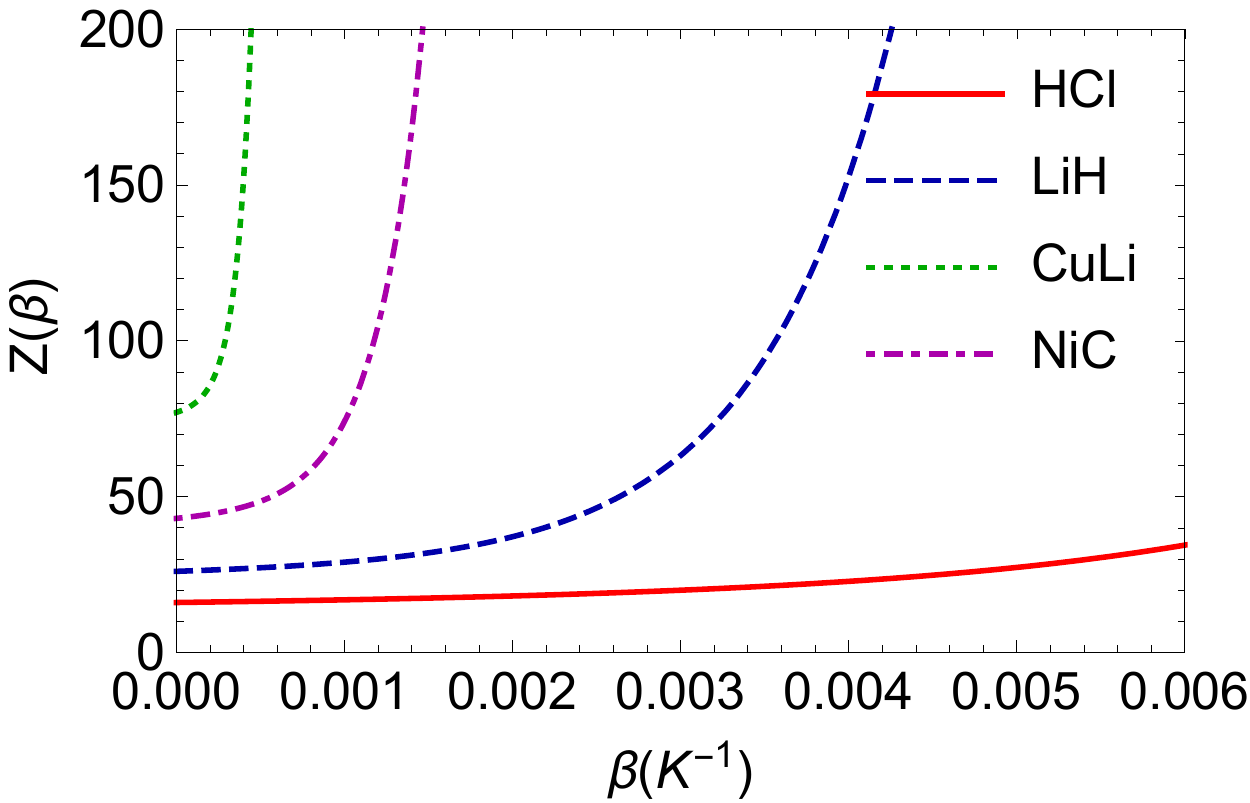}}
	\subfigure[]{\includegraphics[scale=0.54]{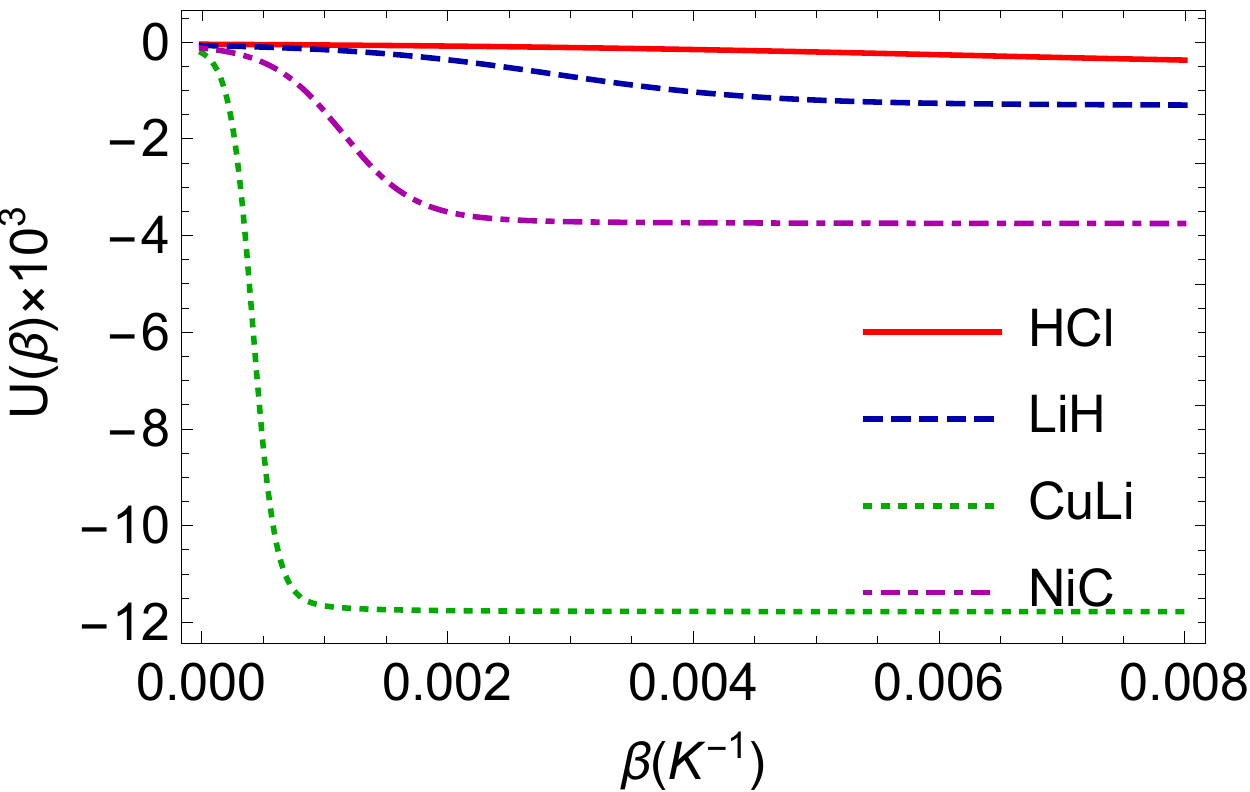}}
	\subfigure[]{\includegraphics[scale=0.54]{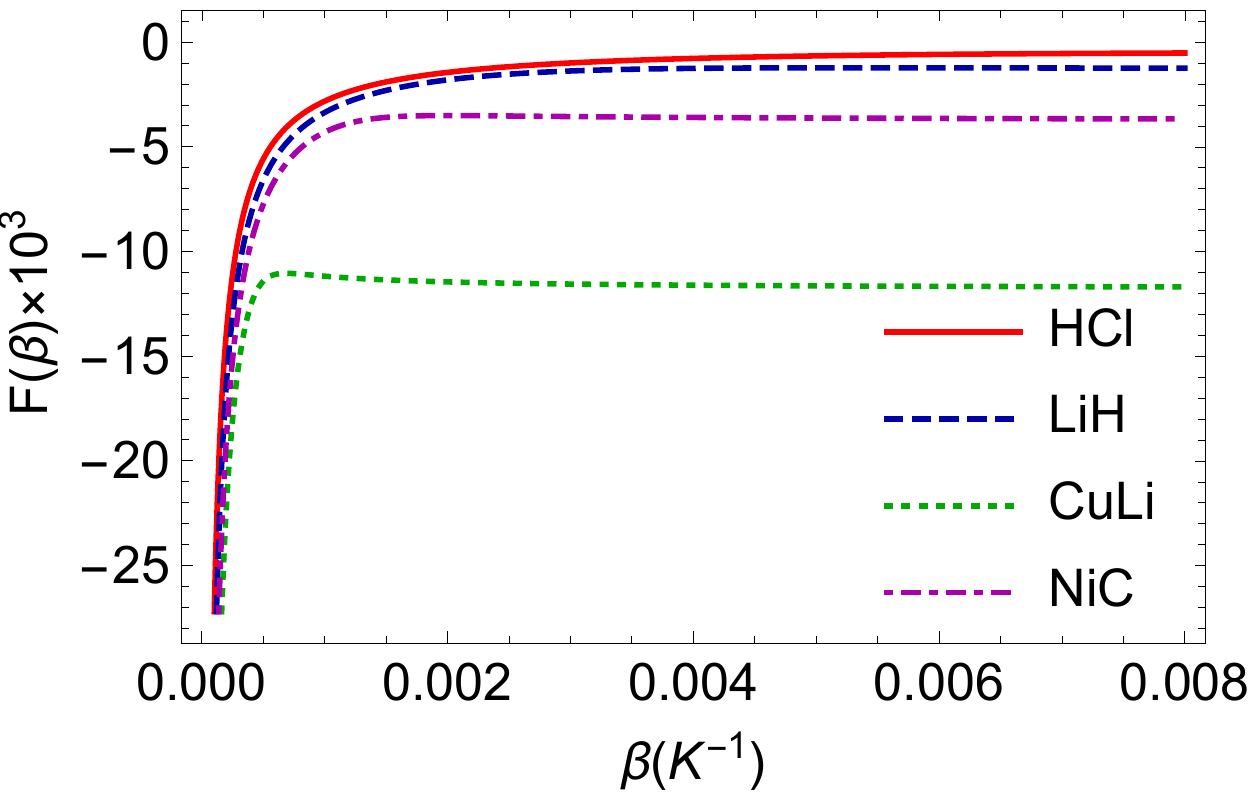}}
	\subfigure[]{\includegraphics[scale=0.54]{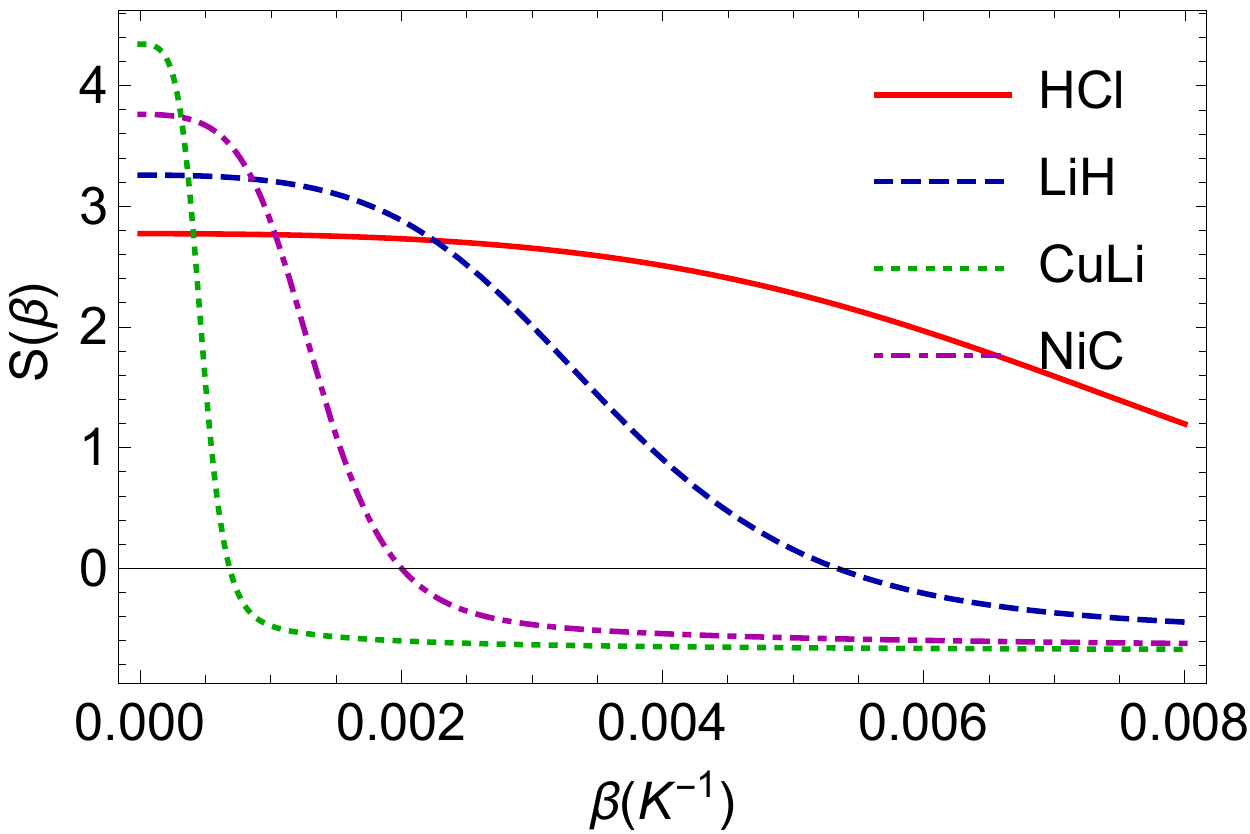}}
	\subfigure[]{\includegraphics[scale=0.54]{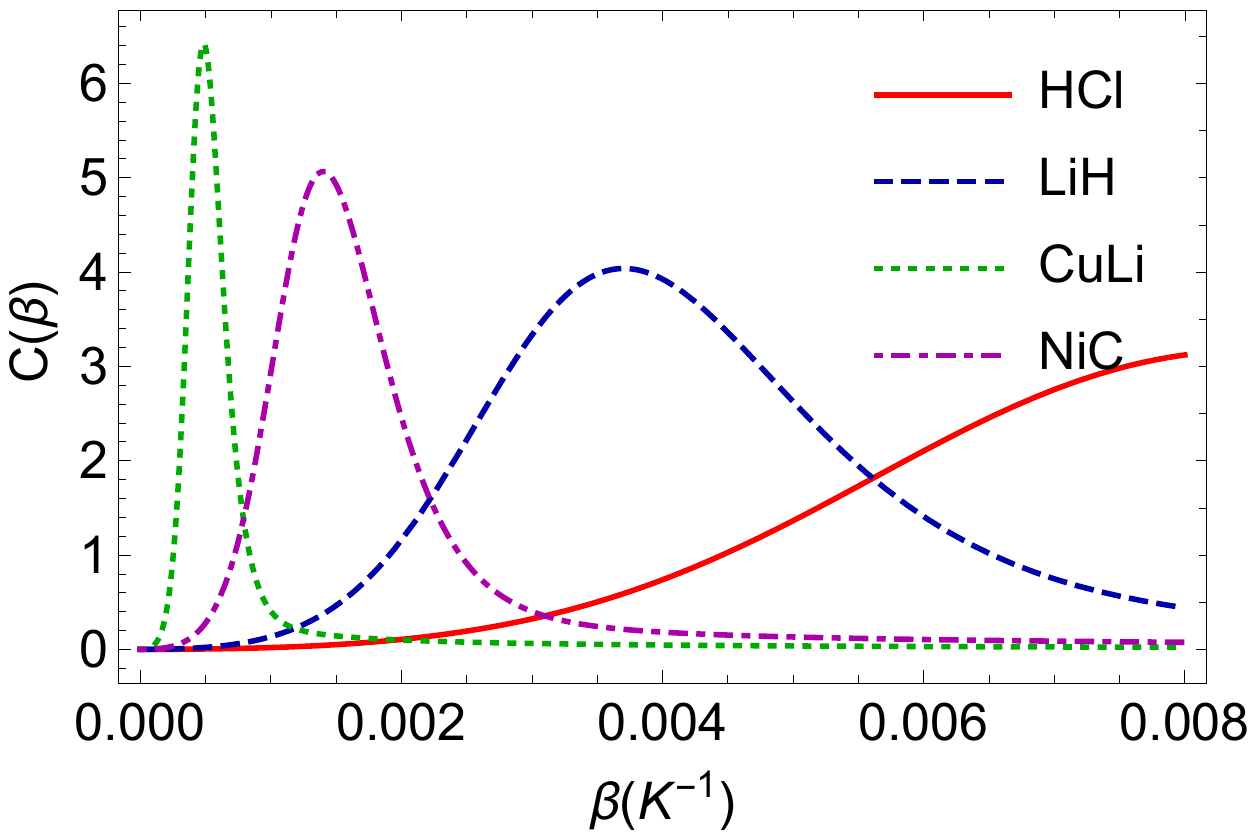}}
     \end{center}
 \vspace{-4mm}
\caption{Energy spectra and thermodynamic quantities as a function of $\beta$ for various diatomic molecules.}\label{fig:Termal_diamol}
\end{figure}

We show the energy spectra for the diatomic molecules $LiH$, $HCl$, $CuLi$ and $NiC$ as a function of angular momentum $l$ in Fig. \ref{fig:Termal_diamol}(a). We observe from this figure that energy spectra of the system increases monotonically with increment in $l$. In Fig. \ref{fig:Termal_diamol}(b-f), we present the $\beta$-dependence of the partition function, mean energy, free energy, entropy, and specific heat capacity for the considered diatomic molecules. It is clear that the beta dependencies here are similar to the behavior observed in Figs.~\ref{fig:Termall} and~\ref{fig:Termal_del} for the general case with arbitrary parameters. For example, the partition function $Z(\beta)$ increases exponentially, while the mean energy and entropy decrease as the $\beta$ increases for each molecule. We obtain the maximum principal quantum number $N$ from Eq.\eqref{Eq:lambda} as $N=15,25,76,42$ for $HCl$, $LiH$, $CuLi$ and $NiC$, respectively. The partition function $Z(\beta)$ increases exponentially faster in diatomic molecules with larger $N$ as well as larger reduced mass $\mu$. The mean and free energies of $CuLi$ is smaller than the others. The decline in entropy $S(\beta)$ with the increment in $\beta$ is fastest in $CuLi$ and slowest in $HCl$ among selected diatomic molecules. The specific heat capacity $C(\beta)$ first peaks and then decreases as the $\beta$ increases for all considered molecules. For large $N$, the height of the peak becomes larger. The width of the peak is the smallest for $CuLi$ and the largest for $HCl$. 

Our results are in good agreement with those obtained in Fig.(2) of Ref. \cite{Ikot23}, where they have predicted the thermal properties for diatomic molecules $CrH$, $CuLi$ and $NiC$ by using Hulthén-type plus Yukawa potential. However, they have used the same $N$ value for all molecules without using Eq. \eqref{Eq:lambda}. 

\section{Conclusions}\label{cr}
In this work, we have first derived any $\ell$ state bound solutions of the KG equation for a novel combined potential: the Eckart plus a class of Yukawa potential. In this regard, we have used the parametric NU method and applied the improved approximation scheme to deal with the centrifugal term. We have derived analytical expressions of energy eigenvalues and normalized wave functions for any quantum states $n_r$ and $\ell$.

It is obvious that the bound-state solutions of the Eckart plus a class of Yukawa potential are more stable compared to the separated cases. The energy eigenvalues are sensitive with regard to the quantum numbers $n_r$ and $\ell$ as well as the parameter $\delta$. Our results show that energy eigenvalues are more bounded at either smaller quantum number $\ell$ or smaller parameter $\delta$. The wave function has $(n_r+1)$ and $(\ell+1)$ nodes. The number of radial nodes is not affected by the position dependence of the strength of combined potential.

We have presented some particular cases, the central and quadratic Yukawa potentials, Eckart potential, Hulthén potential, Kratzer–Fues potential, and Coulomb-like potential, constructed by adjusting the potential parameters. We have derived the energy spectrum equations for these particular cases and showed that they are in agreement with the reports from the previous studies. 

Furthermore, we have derived the analytical expressions of non-relativistic energy spectrum and the thermodynamic quantities, including partition function $Z(\beta)$, mean energy $U(\beta)$, free energy $F(\beta)$, entropy $S(\beta)$, and specific heat capacity $C(\beta)$, for the Eckart plus a class of Yukawa potential at the non-relativistic limit. Then, we have presented the dependence of these quantities on the parameter $\beta$ for various $\ell$ and $\delta$. As detailed in the previous section, the thermodynamic quantities are strongly influenced by the parameters $\beta$ (inverse temperature) and $\delta$. We have predicted the thermal properties for a few diatomic molecules $LiH$, $HCl$, $CuLi$ and $NiC$ through our potential model.

The method used in this study is systematic one, and in many cases, it is one of the most reliable techniques in the research fields. The potential model, which consists of the linear combination of the Eckart and a class of Yukawa potential, could be one of the significant exponential potentials and deserves special attention in many branches of physics, particularly atomic, molecular, nuclear and particle physics.

\section*{Data Availability}
All data generated or analyzed during this study are included in the article.
%


\end{document}